\documentclass{aastex62}

\hypersetup{linkcolor=red,citecolor=blue,filecolor=cyan,urlcolor=magenta}

\usepackage{multirow}

\usepackage{ulem}
\usepackage{lineno}

\usepackage{natbib}
\usepackage{amsmath}
\usepackage{gensymb}

\newcommand\source{OJ~287}
\newcommand\swift{\textit{Swift}}
\newcommand\xrt{\textit{Swift}-XRT}
\newcommand\uvot{\textit{Swift}-UVOT}
\newcommand\fermi{\textit{Fermi}}
\newcommand\lat{\textit{Fermi}-LAT}

\newcommand\gev{\mathrm{~GeV}}

\graphicspath{{./}{Figures/}}

\received{\today}
\revised{TBD}
\accepted{TBD}
\submitjournal{ApJ}

\shorttitle{MWL Study of OJ~287 During 2017}
\shortauthors{VERITAS et al.}

\begin{document}

\title{A multi-wavelength study to decipher the 2017 flare of the blazar OJ 287}



\author[0000-0002-2028-9230]{A.~Acharyya}\affiliation{Department of Physics and Astronomy, University of Alabama, Tuscaloosa, AL 35487, USA}
\author[0000-0002-9021-6192]{C.~B.~Adams}\affiliation{Physics Department, Columbia University, New York, NY 10027, USA}
\author{A.~Archer}\affiliation{Department of Physics and Astronomy, DePauw University, Greencastle, IN 46135-0037, USA}
\author[0000-0002-3886-3739]{P.~Bangale}\affiliation{Department of Physics and Astronomy and the Bartol Research Institute, University of Delaware, Newark, DE 19716, USA}
\author[0000-0002-9675-7328]{J.~T.~Bartkoske}\affiliation{Department of Physics and Astronomy, University of Utah, Salt Lake City, UT 84112, USA}
\author{P.~Batista}\affiliation{DESY, Platanenallee 6, 15738 Zeuthen, Germany}
\author[0000-0003-2098-170X]{W.~Benbow}\affiliation{Center for Astrophysics $|$ Harvard \& Smithsonian, Cambridge, MA 02138, USA}
\author[0000-0002-6208-5244]{A.~Brill}\affiliation{N.A.S.A./Goddard Space-Flight Center, Code 661, Greenbelt, MD 20771, USA}
\author{J.~P.~Caldwell}\affiliation{Aerospace Engineering Department, California Polytechnic State University, San Luis Obispo, CA 93407, USA}
\author{M. Carini}\affiliation{Western Kentucky University, Bowling Green, KY 42101, USA}\author{J.~L.~Christiansen}\affiliation{Physics Department, California Polytechnic State University, San Luis Obispo, CA 94307, USA}
\author{A.~J.~Chromey}\affiliation{Center for Astrophysics $|$ Harvard \& Smithsonian, Cambridge, MA 02138, USA}
\author[0000-0002-1853-863X]{M.~Errando}\affiliation{Department of Physics, Washington University, St. Louis, MO 63130, USA}
\author[0000-0002-5068-7344]{A.~Falcone}\affiliation{Department of Astronomy and Astrophysics, 525 Davey Lab, Pennsylvania State University, University Park, PA 16802, USA}
\author[0000-0001-6674-4238]{Q.~Feng}\affiliation{Department of Physics and Astronomy, University of Utah, Salt Lake City, UT 84112, USA}
\author{J.~P.~Finley}\affiliation{Department of Physics and Astronomy, Purdue University, West Lafayette, IN 47907, USA}
\author[0000-0002-2944-6060]{J.~Foote}\affiliation{Department of Physics and Astronomy and the Bartol Research Institute, University of Delaware, Newark, DE 19716, USA}
\author[0000-0002-1067-8558]{L.~Fortson}\affiliation{School of Physics and Astronomy, University of Minnesota, Minneapolis, MN 55455, USA}
\author[0000-0003-1614-1273]{A.~Furniss}\affiliation{Department of Physics, California State University - East Bay, Hayward, CA 94542, USA}
\author{G.~Gallagher}\affiliation{Department of Physics and Astronomy, Ball State University, Muncie, IN 47306, USA}
\author[0000-0002-0109-4737]{W.~Hanlon}\affiliation{Center for Astrophysics $|$ Harvard \& Smithsonian, Cambridge, MA 02138, USA}
\author[0000-0002-8513-5603]{D.~Hanna}\affiliation{Physics Department, McGill University, Montreal, QC H3A 2T8, Canada}
\author[0000-0003-3878-1677]{O.~Hervet}\affiliation{Santa Cruz Institute for Particle Physics and Department of Physics, University of California, Santa Cruz, CA 95064, USA}
\author[0000-0001-6951-2299]{C.~E.~Hinrichs}\affiliation{Center for Astrophysics $|$ Harvard \& Smithsonian, Cambridge, MA 02138, USA and Department of Physics and Astronomy, Dartmouth College, 6127 Wilder Laboratory, Hanover, NH 03755 USA}
\author{J.~Hoang}\affiliation{Santa Cruz Institute for Particle Physics and Department of Physics, University of California, Santa Cruz, CA 95064, USA}
\author[0000-0002-6833-0474]{J.~Holder}\affiliation{Department of Physics and Astronomy and the Bartol Research Institute, University of Delaware, Newark, DE 19716, USA}
\author[0000-0002-1432-7771]{T.~B.~Humensky}\affiliation{Department of Physics, University of Maryland, College Park, MD, USA and NASA GSFC, Greenbelt, MD 20771, USA}
\author[0000-0002-1089-1754]{W.~Jin}\affiliation{Department of Physics and Astronomy, University of California, Los Angeles, CA 90095, USA}
\author[0009-0008-2688-0815]{M.~N.~Johnson}\affiliation{Santa Cruz Institute for Particle Physics and Department of Physics, University of California, Santa Cruz, CA 95064, USA}
\author[0000-0002-3638-0637]{P.~Kaaret}\affiliation{Department of Physics and Astronomy, University of Iowa, Van Allen Hall, Iowa City, IA 52242, USA}
\author{M.~Kertzman}\affiliation{Department of Physics and Astronomy, DePauw University, Greencastle, IN 46135-0037, USA}
\author{M.~Kherlakian}\affiliation{DESY, Platanenallee 6, 15738 Zeuthen, Germany}
\author[0000-0003-4785-0101]{D.~Kieda}\affiliation{Department of Physics and Astronomy, University of Utah, Salt Lake City, UT 84112, USA}
\author[0000-0002-4260-9186]{T.~K.~Kleiner}\affiliation{DESY, Platanenallee 6, 15738 Zeuthen, Germany}
\author[0000-0002-4289-7106]{N.~Korzoun}\affiliation{Department of Physics and Astronomy and the Bartol Research Institute, University of Delaware, Newark, DE 19716, USA}
\author{F.~Krennrich}\affiliation{Department of Physics and Astronomy, Iowa State University, Ames, IA 50011, USA}
\author[0000-0002-5167-1221]{S.~Kumar}\affiliation{Department of Physics, University of Maryland, College Park, MD, USA }
\author[0000-0003-4641-4201]{M.~J.~Lang}\affiliation{School of Natural Sciences, University of Galway, University Road, Galway, H91 TK33, Ireland}
\author[0000-0003-3802-1619]{M.~Lundy}\affiliation{Physics Department, McGill University, Montreal, QC H3A 2T8, Canada}
\author[0000-0001-9868-4700]{G.~Maier}\affiliation{DESY, Platanenallee 6, 15738 Zeuthen, Germany}
\author{C.~E~McGrath}\affiliation{School of Physics, University College Dublin, Belfield, Dublin 4, Ireland}
\author[0000-0001-7106-8502]{M.~J.~Millard}\affiliation{Department of Physics and Astronomy, University of Iowa, Van Allen Hall, Iowa City, IA 52242, USA}
\author{J.~Millis}\affiliation{Department of Physics and Astronomy, Ball State University, Muncie, IN 47306, USA}
\author[0000-0001-5937-446X]{C.~L.~Mooney}\affiliation{Department of Physics and Astronomy and the Bartol Research Institute, University of Delaware, Newark, DE 19716, USA}
\author[0000-0002-1499-2667]{P.~Moriarty}\affiliation{School of Natural Sciences, University of Galway, University Road, Galway, H91 TK33, Ireland}
\author[0000-0002-3223-0754]{R.~Mukherjee}\affiliation{Department of Physics and Astronomy, Barnard College, Columbia University, NY 10027, USA}
\author[0000-0002-9296-2981]{S.~O'Brien}\affiliation{Physics Department, McGill University, Montreal, QC H3A 2T8}
\affiliation{Trottier Space Institute, McGill University, Montreal, QC H3A 2A7, Canada}
\affiliation{Arthur B. McDonald Canadian Astroparticle Physics Research Institute, 64 Bader Lane, Queen's University, Kingston, ON Canada, K7L 3N6}
\author[0000-0002-4837-5253]{R.~A.~Ong}\affiliation{Department of Physics and Astronomy, University of California, Los Angeles, CA 90095, USA}
\author[0000-0001-7861-1707]{M.~Pohl}\affiliation{Institute of Physics and Astronomy, University of Potsdam, 14476 Potsdam-Golm, Germany and DESY, Platanenallee 6, 15738 Zeuthen, Germany}
\author[0000-0002-0529-1973]{E.~Pueschel}\affiliation{Fakult\"at f\"ur Physik \& Astronomie, Ruhr-Universit\"at Bochum, D-44780 Bochum, Germany}
\author[0000-0002-4855-2694]{J.~Quinn}\affiliation{School of Physics, University College Dublin, Belfield, Dublin 4, Ireland}
\author{P.~L.~Rabinowitz}\affiliation{Department of Physics, Washington University, St. Louis, MO 63130, USA}
\author[0000-0002-5351-3323]{K.~Ragan}\affiliation{Physics Department, McGill University, Montreal, QC H3A 2T8, Canada}
\author{P.~T.~Reynolds}\affiliation{Department of Physical Sciences, Munster Technological University, Bishopstown, Cork, T12 P928, Ireland}
\author[0000-0002-7523-7366]{D.~Ribeiro}\affiliation{School of Physics and Astronomy, University of Minnesota, Minneapolis, MN 55455, USA}
\author{E.~Roache}\affiliation{Center for Astrophysics $|$ Harvard \& Smithsonian, Cambridge, MA 02138, USA}
\author[0000-0001-6662-5925]{J.~L.~Ryan}\affiliation{Department of Physics and Astronomy, University of California, Los Angeles, CA 90095, USA}
\author[0000-0003-1387-8915]{I.~Sadeh}\affiliation{DESY, Platanenallee 6, 15738 Zeuthen, Germany}
\author{A.~C.~Sadun}\affiliation{Department of Physics, University of Colorado Denver, Campus Box 157, P.O. Box 173364, Denver CO 80217, USA}
\author[0000-0002-3171-5039]{L.~Saha}\affiliation{Center for Astrophysics $|$ Harvard \& Smithsonian, Cambridge, MA 02138, USA}
\author{M.~Santander}\affiliation{Department of Physics and Astronomy, University of Alabama, Tuscaloosa, AL 35487, USA}
\author{G.~H.~Sembroski}\affiliation{Department of Physics and Astronomy, Purdue University, West Lafayette, IN 47907, USA}
\author{K.~Shahinyan}\affiliation{School of Physics and Astronomy, University of Minnesota, Minneapolis, MN 55455, USA}
\author[0000-0002-9856-989X]{R.~Shang}\affiliation{Department of Physics and Astronomy, Barnard College, Columbia University, NY 10027, USA}
\author[0000-0003-3407-9936]{M.~Splettstoesser}\affiliation{Santa Cruz Institute for Particle Physics and Department of Physics, University of California, Santa Cruz, CA 95064, USA}
\author[0000-0002-9852-2469]{D.~Tak}\affiliation{SNU Astronomy Research Center, Seoul National University, Seoul 08826, Republic of Korea.}
\author{A.~K.~Talluri}\affiliation{School of Physics and Astronomy, University of Minnesota, Minneapolis, MN 55455, USA}
\author{J.~V.~Tucci}\affiliation{Department of Physics, Indiana University-Purdue University Indianapolis, Indianapolis, IN 46202, USA}
\author[0000-0003-2740-9714]{D.~A.~Williams}\affiliation{Santa Cruz Institute for Particle Physics and Department of Physics, University of California, Santa Cruz, CA 95064, USA}
\author[0000-0002-2730-2733]{S.~L.~Wong}\affiliation{Physics Department, McGill University, Montreal, QC H3A 2T8, Canada}

\correspondingauthor{Stephan O'Brien}
\email{stephan.obrien@mcgill.ca}
\correspondingauthor{Olivier Hervet}
\email{ohervet@ucsc.edu}

\collaboration{VERITAS Collaboration}

\author{S. G. Jorstad}
\affiliation{Institute for Astrophysical Research, Boston University, 725 Commonwealth Avenue, Boston, MA 02215, USA}

\author{R. Lico}
\affiliation{Instituto de Astrofísica de Andalucía, IAA-CSIC, Apdo. 3004, 18080 Granada, Spain}
\affiliation{INAF Istituto di Radioastronomia, via Gobetti 101, 40129 Bologna, Italy}

\author{P. Lusen}
\affiliation{Santa Cruz Institute for Particle Physics and Department of Physics, University of California, Santa Cruz, CA 95064, USA}

\author{A. P. Marscher}
\affiliation{Institute for Astrophysical Research, Boston University, 725 Commonwealth Avenue, Boston, MA 02215, USA}

\begin{abstract}

In February 2017, the blazar OJ~287 underwent a period of intense multiwavelength activity. 
It reached a new historic peak in the soft X-ray (0.3-10 keV) band, as measured by \xrt. 
This event coincides with a very-high-energy (VHE) $\gamma$-ray outburst that led VERITAS to detect emission above 100 GeV, with a detection significance of $10\sigma$ (from 2016 December 9 to 2017 March 31).
The time-averaged VHE $\gamma$-ray spectrum was consistent with a soft power law ($\Gamma = -3.81 \pm 0.26$) and an integral flux corresponding to $\sim2.4\%$ that of the Crab Nebula above the same energy.
Contemporaneous data from multiple instruments across the electromagnetic spectrum reveal complex flaring behavior, primarily in the soft X-ray and VHE bands. 
To investigate the possible origin of such an event, our study focuses on three distinct activity states: before, during, and after the February 2017 peak.
The spectral energy distributions during these periods suggest the presence of at least two non-thermal emission zones, with the more compact one responsible for the observed flare.
Broadband modeling results and observations of a new radio knot in the jet of OJ~287 in 2017 are consistent with a flare originating from a strong recollimation shock outside the radio core.

\end{abstract}

\keywords{ galaxies: BL Lacertae objects: OJ~287 --- galaxies: jets --- gamma rays: galaxies --- X-rays: galaxies }

\section{Introduction} \label{sec:intro}

At the time of writing, there are 88 active galactic nuclei (AGN) detected at very high energies (VHE, $E > 100 ~\mathrm{ GeV}$) \citep[see \href{http://tevcat2.uchicago.edu/}{TeVCat},][]{TeVCat2008}. 
The population of VHE AGN is dominated by blazars ($\sim$ 90\%), with the remaining detected AGN being either radio galaxies or AGN with uncertain classifications. 
Blazars are divided into two classes, BL Lacertae objects (BL Lacs) and flat spectrum radio quasars (FSRQs). 

The spectral energy distribution (SED) of blazars is characterized by a double-peaked structure.
The lower-frequency peak, typically found at radio-X-ray frequencies, is due to synchrotron radiation of relativistic electrons traveling within the jet.
The higher-frequency peak, typically located in the X-ray-$\gamma$-ray regime, is generally attributed to inverse-Compton (IC) scattering of low-energy photons.
The origin of these low-energy photons may be the same photons produced by the synchrotron process (synchrotron-self Compton, SSC, see, for example, \cite{1996A&AS..120C.503G,1998ApJ...509..608T}) or due to an external region (external-Compton, EC, see, for example, \cite{1993ApJ...416..458D}), such as the broad line region (BLR) or a dusty torus. 
Hadronic and lepto-hadronic models have also been invoked in modeling blazars \citep[see, for example,][]{2012arXiv1205.0539B,2015MNRAS.448..910C}.
See \citet{2020Galax...8...72C} and references therein for a review of blazar emission models.

\par The BL Lac classification is further broken down by the frequency of the synchrotron peak, low- (LBL, $\nu_{sync} < 10^{14}$ Hz), intermediate- (IBL, $\nu_{sync}\sim 10^{14}\mathrm{-}10^{15}$ Hz) or high-frequency-peaked BL Lacs (HBL $\nu_{sync} > 10^{15}$ Hz) \citep[see, for example,][]{Padovani1995, Nieppola2006}.
HBLs are generally well described by a one-zone SSC model, while LBLs and IBLs usually are better described by significant  EC components.
HBLs are the dominant class of BL Lacs detected at VHE energies, with LBLs and IBLs constituting $\sim$15\% of the detected BL Lacs.
LBLs and IBLs are typically only detected during periods of enhanced VHE and multiwavelength (MWL) activity, with simple one-zone SSC models struggling to adequately explain the broadband SED \citep[see, for example, BL Lac and Ap Librae,][ respectively.]{2010A&A...524A..43R, Hervet2015}.

\par \source\ \citep[R.A.: 08h~54\arcmin\ 48.8749\arcsec\, Dec: +20h 06\arcmin\ 30.641\arcsec\ (J2000), ][]{Johnston1995} is a blazar located at a redshift of z = 0.306 \citep{Nilsson2010}.

OJ~287 belongs to the LBL subclass of blazars, with a synchrotron peak (in $\nu F_{\nu}$ representation) at 
$\log_{10}(\nu\mathrm{~[Hz]}) = 13.24$ \citep{4LACDR2_2020}. 
 \source\ is a remarkably well-studied object, with optical observations dating back to 1890 \citep{Sillanpaa1988}.
These optical observations have revealed a quasi-periodic behavior, with regular outbursts occurring on an approximately 12 year cycle.
To explain this behavior, current models invoke a precessing binary supermassive black hole (SMBH) system at the center of the AGN \citep[see, for example,][]{Sillanpaa1988, Sundelius_1997, Valtonen2011, Hudec2013} or helical jet models \citep[see, for example,][]{Valtonen2013, Britzen2018}.

The binary SMBH models typically invoke a primary and a secondary black hole with mass of $1.8\times 10^{10} \mathrm{~M_\odot}$ and $1.3\times 10^{8} \mathrm{M_\odot}$ respectively. and precession rate of $\left(37.5 \mathrm{-} 39.1\right)^\circ\mathrm{orbit^{-1}}$ \citep{2007ApJ...659.1074V, Valtonen2011}.
These models have successfully predicted optical outbursts originating from a disk-crossing event \citep{Valtonen2011}, with the recent event \citep{Valtonen:2016dy} in July 2019\textcolor{red}{,} occurring within 4 hours of the predicted flare \citep{2020ApJ...894L...1L}.
\citet{Valtonen:2016dy} observed the optical flare on the 5th of December 2015 as part of a multiwavelength campaign spanning optical to X-ray energies.
Optical R-band observations taken during this campaign show good agreement with the thermal bremsstrahlung component of an optical outburst model \citep[see,][and references therein]{2012MNRAS.427...77V}, with flux levels in excess of the model, consistent with variations observed in  optical polarization and X-ray measurements during this period, suggesting a synchrotron origin for the excess emission.
The X-ray flux level observed during this campaign was comparable to that observed during a previous monitoring campaign ($\sim 4\times10^{-12}\mathrm{erg~cm^{-2}~s^{-1}}$ ) \citep[][]{Valtonen:2016dy, 2015ATel.7056....1E}, suggesting that the X-ray emission is dominated by jet emission rather than the disk-crossing event.

\source\ has been detected at high energies (HE, 100 MeV - 100 GeV) by \lat.
The 4FGL \citep[DR1,][]{4FGL_2020} energy spectrum is described by a log-parabola model, with the inverse-Compton peak occurring in the MeV to sub-GeV energy range.
A conservative power-law extrapolation of the 4FGL spectrum into the VHE regime suggests a VHE flux $<0.5\%$ Crab\footnote{A flux unit defined as the VHE flux of the Crab Nebula \citep{1998ApJ...503..744H}, a bright and steady source at VHE.}, indicating that in its quiescent state, a detection of \source\ would not be feasible by current-generation ground-based gamma-ray instruments without significant ($>$ 50 hours) exposure.
\source\ is highly variable at HE, with a 4FGL variability index ($TS_{var}$) of 611.67. A 4FGL variability index of $TS_{var} > 18.48$ suggests, at the 99\% confidence level, that the source is inconsistent with a constant-flux model.
Indeed, EGRET observations already suggested HE variability \citep{Egret_OJ287_1996A&AS}, with a $4.3\sigma$ excess of events $>1 \mathrm{ ~GeV}$ observed coincident with optical flaring.

\source\ has previously been observed in VHE $\gamma$-rays by the VERITAS and MAGIC Collaborations.
VERITAS observed \source\ during a period of expected enhanced activity between December 4th, 2007 and January 1st, 2008 \citep[see, for example,][]{2007ApJ...659.1074V}, resulting in a non-detection.
\citet{Archambault:2016ha} combined these observations with ones taken during the 2010-2011 season, again resulting in a non-detection, and determined a 99\% confidence level (C.L.) upper limit on the integral flux ($E > 182 \gev$) at 2.6\% Crab \citep{Archambault:2016ha}.
As part of a multiwavelength campaign \citep{Seta2009}, the MAGIC Collaboration observed \source\ during two periods in 2007, April 10th to 13th and 2007, November 7th to 9th. 
The data from these observations both resulted in a non-detection.
\citet{Seta2009} determined 95\% C.L. upper limits on the integral flux at ($E > 145 \gev$) 3.3\% and ($E > 150 \gev$) 1.7\% Crab for the respective periods.

Between late 2016 and mid 2017, \source\ underwent a period of enhanced multiwavelength activity.
In December 2016, VERITAS began a monitoring program on \source\ based on elevated X-ray count rates  reported by \href{https://www.swift.psu.edu/monitoring/}{the Swift-XRT Monitoring of Fermi-LAT Sources of Interest} website \citep{swift-mon2013} \citep[see also, ][]{2016ATel.9629....1G,2016ATel.9709....1V}.
In response to a further increase in the X-ray count rates, VERITAS initiated target of opportunity observations on \source\ from 2017 February  1 to  2017 February 4, resulting in a $>5\sigma$ detection \citep{2017ATel10051....1M}.
Preliminary analysis results of the VERITAS data were reported at the 2017 International Cosmic Ray Conference \citep{OBrien_2017}.


In this work, the multiwavelength data from radio to VHE are presented.
In Section \ref{obs}, the multiwavelength observations, spanning from optical to VHE, taken between 2016 December 1  and 2017 April 15 (57723 - 57858 MJD) are discussed, and the key results are presented.
In Section \ref{temporal}, the temporal flux properties are discussed, and the correlation between different energy bands is studied in Section \ref{cor}.
In Section \ref{sed_modeling}, the time-averaged SEDs obtained for three periods are presented, and modeling of the broadband SED is performed. A general discussion of our results and their context with respect to other recent studies are presented in Section \ref{Section::Discussion}. We summarize our conclusions in Section \ref{Section::Conclusion}.
This paper uses a flat $\Lambda$CDM cosmology, with $H_0$ = 71 km s$^{-1}$ Mpc$^{-1}$, $\Omega M = 0.286$, and $\Omega \Lambda = 0.714$.


\section{Observations}\label{obs}
\subsection{VERITAS}\label{obs:ver}
The Very Energetic Radiation Imaging Telescope Array System \citep[VERITAS, ][]{holder2006first}, is an array of four 12m imaging atmospheric-Cherenkov telescopes (IACTs) located at the Fred Lawrence Whipple Observatory in southern Arizona, USA 
(31$^\circ$ 40\arcmin~30\arcsec N, 110$^{\circ}$ 57\arcmin~07\arcsec W,  1.3km above sea level).
In its current configuration, VERITAS can accurately reconstruct $\gamma$-ray events in the $\sim 100~\mathrm{GeV}$ and $>30~\mathrm{TeV}$ energy range.
VERITAS can detect a source with a flux of 1\% Crab at 5$\sigma$ in $\sim$25 hours of observation. 
For a detailed discussion of the performance of VERITAS see \cite{park2015performance, 2022A&A...658A..83A}.

\par VERITAS observations of \source ~between 2016  December 9  (57731 MJD) and 2017 March 30 (57842 MJD) resulted in a total of 57 hours of quality-selected and deadtime-corrected exposure. 
The VERITAS data were analyzed using standard analysis techniques  \citep{acciari2008veritas} and crosschecked with two independently developed and maintained analysis packages \citep{EDMaier2017,VEGASCogan2007,christiansen2017}. Excellent agreement was found between the two packages. 
Data were corrected for PMT gain \citep{2022NIMPA102766235H} and throughput losses \citep{2021APh...12802556A} using the methods described by \citet{2022A&A...658A..83A}.
Gamma-hadron separation was performed using a set of boosted decision tree (BDT) cuts optimized and verified, a priori, for analysis of soft-spectrum sources \citep[see][]{Krause2017}.
This resulted in the detection of 2210 on-source and 10327 off-source $\gamma$-ray-like events, with an on/off normalization of 1/6.
Hence, OJ~287 is detected with an excess significance \citep[Equation 17 of][]{Li1983} of 10.4$\sigma$.
To determine the best-fit source location, a 2D symmetric Gaussian was fitted to the excess counts map. 
The best-fit source location is determined to be (J2000) R.A: 08h 54$\arcmin$ 50.3$\arcsec$ $\pm$ (2.0$\arcsec$)$_{stat}$,  Dec: 20$^\circ$ 06$\arcmin$ 25.4$\arcsec$ $\pm$ (29.2$\arcsec$)$_{stat}$, which is consistent with the radio location of \source\ \citep{Johnston1995}.
The 68\% containment of the best-fit location is $0.119^\circ \pm 0.008^\circ$, which is consistent with observations of a point source convolved  with the VERITAS PSF of $0.1^\circ$. 
The systematic uncertainty on the VERITAS pointing accuracy is $< 25\arcsec$.
\source~ is therefore assigned the VERITAS catalog name VER~J0854+201.

\par Spectral analysis was performed by applying a forward-folding binned-likelihood analysis method described by \citet{Piron2001}.
The total time-averaged differential energy spectrum between 110 GeV and 630 GeV is fitted ($\chi^2/NDF = 1.45/3 = 0.48$) by a power law of the form:
\begin{equation}
\frac{dN}{dE} = \left( 4.33 \pm 0.43\right)\times 10^{-12} \left( \frac{E}{ 0.2 \mathrm{~TeV}} \right)^{-3.81 \pm 0.26} [\mathrm{cm^{-2} s ^{-1} TeV^{-1}}].    
\end{equation}

\begin{figure}
  \plotone{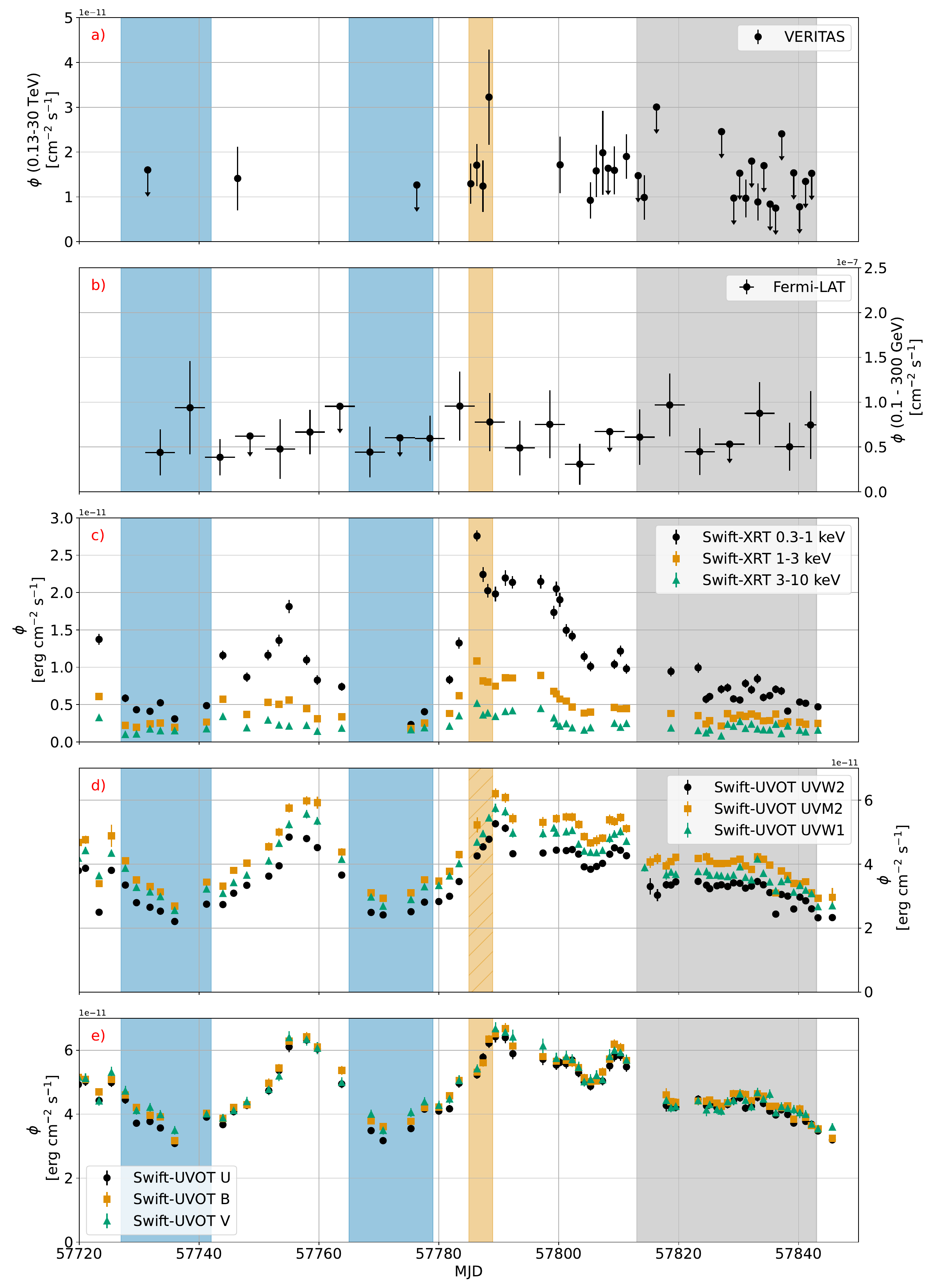}
  \caption{
  			 Multiwavelength light curves of \source. Panel (a) shows the VHE flux above 130 GeV obtained by VERITAS. Panel (b) shows the flux between 0.1 GeV and 300 GeV obtained by \lat. 
   			Panel (c)  shows the X-ray flux broken down into soft (0.3-1 keV), medium (1-3 keV) and hard (3-10 keV) energy bands. 
   			Panels (d) and (e) show the UV and optical observations taken by \uvot\ in the {\it UVW1}, {\it UVM2} and {\it UVW2}, and {\it U}, {\it B} and {\it V} bands respectively.
   			The shaded regions in blue, orange, and grey, correspond to the studied periods Low, Flare, and Post-Flare, respectively, defined in Section \ref{obs:ver}.
            \label{fig:MWL_LC}
        }
\end{figure}

A log-parabola model was also applied, however no significant curvature was found. 
The total time-averaged integral flux was determined to be $\phi_{(E > 130 \mathrm{~GeV})} = \left(1.04\pm0.10\right)\times 10 ^{-11} \mathrm{~cm^{-2}~s^{-1}}$ or $ \sim 2.4 \%$ Crab.
This is inconsistent with the flux upper limits set by \cite{Seta2009}, suggesting that this detection corresponds to a period of enhanced VHE activity rather than steady emission.
The nightly-binned integral flux is shown in panel (a) of Figure~\ref{fig:MWL_LC}. Upper limits at the  95\% C.L. are obtained for observations with an excess significance $< 2\sigma$, using the bound likelihood method described by \citet{rolke2005}.

\par 
The variability index \citep[see, for example,][]{2FGL2012} of the nightly-binned VERITAS data 
is found to be $TS_{var} = 30.14$. 
This corresponds to a $\chi^2/NDF$ of $30.14 / 31 = 0.97$, which has a $\chi^2$-probability of $p = 0.51$, suggesting a nightly-binned flux consistent with a constant-flux model.
The validity of the $TS_{var}$ as an equivalent $\chi^2$-statistic was verified by Monte Carlo simulations of VERITAS observations taken under similar conditions as the observations reported here. 
The lack of detected VHE variability is not surprising given that \source\ is not consistently detected on nightly timescales. 
Due to the long integration times required to obtain a significant detection, it is more relevant to discuss long-term flux trends.

\par To analyze the temporal evolution of the VHE flux and spectrum, we define three periods of multiwavelength activity. 
We defined a ``Low'' state comprising the periods 57727 to 57742 MJD (2016 December 5 to 2016 December 20)and 57765 to 57779 MJD (2017 January 12 to 2017 January 26), during which there was no enhanced X-ray to optical activity (see Sections \ref{obs:swift} and \ref{obs:opt}). 
The ``Flare'' state is defined as 57785 to 57789 MJD (2017 February 1 to 2017 February 5) and corresponds to the peak of the soft X-ray flux and the VHE detection.
Finally, a ``Post-Flare'' state is defined as 57813 to 57843 MJD (2017 March 1 to 2017 March 31) and corresponds to a period of decreasing X-ray flux with some smaller-scale flux variability. 
A breakdown of these periods with associated VERITAS results is shown in Table \ref{tab:VerResults}. The ``Low'', ``Flare'' and ``Post-Flare'' states are shown as shaded blue, orange and gray regions. respectively, in Figures \ref{fig:MWL_LC}, \ref{fig:optlight} and \ref{fig:optpolar}.

\startlongtable
\begin{deluxetable}{c c c c c c c } 

\tablecaption{
		      Summary of VERITAS results. Columns 1 and 2 show the definition of each analysis period. 
			  Column 3 shows the deadtime-corrected exposure. 
    		  Column 4 shows the excess significance for the period. 
    		  Columns 5 and 6 show the integral flux ($>130$ GeV) in units of $\mathrm{cm^{-2}~s^{-1}}$ and Crab, respectively, with the 95\% C.L. upper limit shown in parenthesis. 
    		  Column 7 shows the best-fit spectral index for the total exposure.
		      \label{tab:VerResults}   
}
\tablehead{
\colhead{Period} & \colhead{Date} & \colhead{Live Time} & \colhead{Significance} & \colhead{Integral Flux} & \colhead{Integral Flux}  & \colhead{Spectral Index} \\ 
 & \colhead{(MJD)} & \colhead{(Hours)} & \colhead{($\sigma$)} & \colhead{$\times10^{-11}$($\mathrm{cm^{-2}~s^{-1}}$)} & \colhead{(\% Crab)}
}
\colnumbers
\startdata
   \multirow{2}{*}{Low State} & 57727 - 57742 & \multirow{2}{*}{3.5} & \multirow{2}{*}{1.3} & \multirow{2}{*}{0.46 $\pm$ 0.40 ($<$1.15)} & \multirow{2}{*}{ 1.07 $\pm$ 0.95 ($<$2.69) }& \multirow{2}{*}{-3.81} \\  
    & 57765 - 57779 & \\
   Flare State & 57785 - 57789 & 10.2 & 6.3 & 1.78 $\pm$ 0.28 ($<$2.26) & 4.18 $\pm$ 0.67 ($<$5.30) & -3.81 \\
   Post-Flare State & 57813 - 57843 & 26.8  & 4.4 & 0.63 $\pm$ 0.14 ($<$0.86)& 1.49 $\pm$ 0.32 ($<$2.02) & -3.81 \\
   \hline
   Total Exposure & 57727 - 57843& 57.0  & 10.4 & 1.04 $\pm$ 0.10 ($<$1.21) & 2.43 $\pm$ 0.24 ($<$2.83) & -3.81 $\pm$ 0.27 \\
\enddata
\tablecomments{
           The total time-averaged best-fit spectral index is assumed when calculating the integral flux for each period.
			  }
\end{deluxetable}

The significance of the change in flux can be estimated by:
\begin{equation}
    \label{eqn:fluxChangeSigma}
    \sigma_{i,j} = \frac{ |\phi_i - \phi_j|}{\sqrt{\Delta\phi_i^2 + \Delta\phi_j^2 } },
\end{equation}
 where $\phi_{i}$ and $\Delta\phi^2_{i}$ are the integral flux and associated squared uncertainty for the $i^{th}$ observation.
From the ``Low'' to ``Flare'' state, the VHE flux increases by a factor of $\sim4$ with a $2.7\sigma$ significance on the increase.
The VHE flux then decreased between the  ``Flare'' and ``Post-Flare'' states by a factor of $\sim3$ with a $3.7\sigma$ significance. 
The ``Low'' and ``Post-Flare'' states show a consistent integral flux.
For each of the defined periods, differential spectral points are obtained by reapplying the spectral fit to each energy bin, with the spectral index frozen to the best-fit value. 

\subsection{\lat}\label{obs:lat}

 The \fermi\ Large Area Telescope \citep[\lat,][]{Atwood2009} is a pair-conversion $\gamma$-ray telescope sensitive to $\gamma$-rays with energies from about 20 MeV to $>$300 GeV.
Data taken by \lat\ between 57731 MJD and 57844 MJD were analyzed using the {\it Science Tools} package (\texttt{fermitools}  v2.2.0) using ``Pass 8'' (P8R3) instrument response functions \citep{Atwood2013}, using the \texttt{fermipy} (v1.2.0) \lat\ analysis suite \citep{Wood:2017TJ}.
``\texttt{Source}'' class events (evclass=128) converting in both the front and back of the instrument (evtype=3) were analyzed. 
Events with energies between 0.1 GeV and 300 GeV were selected.
Events within a 15$^\circ$ of the source were considered, with a zenith cut of 90$^\circ$ applied to remove contamination due to the Earth's limb.

\par A binned-likelihood analysis \citep[see, for example,][]{1996ApJ...461..396M} was applied to events passing the above criteria. 
Different event types (4, 8, 16, 32) were used in a summed-likelihood analysis with the appropriate IRFs  considered for each event class.
A source model was constructed that included all known sources in the 4th \fermi\ Point Source catalog \citep[4FGL-DR1,][]{4FGL_2020} within the region of interest.
In the minimization process, sources within 5$^\circ$ of \source\ and sources with a TS $>$ 10 had their spectral shape parameters fixed to their 4FGL values and their spectral normalization allowed to vary. 
The spectral shape and normalization parameters are free to vary for sources with a TS $>$ 50.

\par The light curve was obtained by binning the data in five-day time bins. 
This is shown in Panel (b) of Figure~\ref{fig:MWL_LC}.
A power-law fit was applied to each of the time bins with the spectral index fixed to $-2.0$.
For bins with a test statistic (TS) less than 9, 95\% C.L. upper limits were obtained.

\par The Fermi-LAT light curve (100 MeV - 300 GeV) shows remarkable stability during the campaign, with frequent detections on a 5-day time scale.
To provide a comparable flux evolution study, the HE analysis was performed across the same time bins as the VHE observations. 
Details of this analysis are reported in Table \ref{tab:LATResults}.
The HE spectra and flux states during each period are consistent with constant emission, within statistical uncertainties.

\startlongtable
\begin{deluxetable}{c c c c} 
\tablecaption{
		      Summary of \lat\ results. Column 1 shows the analysis period as defined in Table \ref{tab:VerResults}. 
			  Column 2 shows Test Statistic values. 
    		  Columns 3 and 4 show the best-fit time-averaged integral fluxes and power-law spectral, respectively. 
		      \label{tab:LATResults}   
}
\tablehead{
\colhead{Period}  & \colhead{Test Statistic} & \colhead{Integral Flux}  & \colhead{Spectral Index} \\ 
  & & \colhead{$\left( 0.1 \mathrm{~GeV} \leq E \leq 300 \mathrm{~GeV}\right)$} & \colhead{}\\
  & \colhead{(TS)} & \colhead{$\times10^{-8}$($\mathrm{cm^{-2}~s^{-1}}$)} & \colhead{}
}
\colnumbers
\startdata
   Low State & 57.78 & $\left(3.88 \pm 1.56\right)$  & $\left( -1.98 \pm 0.20\right)$ \\  
   Flare State & 36.19 & $\left(7.83 \pm 3.43\right)$ & $\left(-1.89 \pm 0.27\right)$ \\
   Post Flare State & 188.16 & $\left(7.03 \pm 1.57\right)$ &  $\left(-1.94 \pm 0.12\right)$ \\
   \hline
   Total Exposure & 531.45 & $\left(6.08 \pm 0.81\right)$ &  $\left(-1.94 \pm 0.07\right)$
\enddata
\end{deluxetable}

\subsection{\swift}\label{obs:swift}

\subsubsection{\xrt}\label{sec:swift:xrt}

On board the Neil Gehrels Swift Observatory (\swift) \citep[][]{2004ApJ...611.1005G}, the X-ray Telescope \citep[\xrt,][]{Burrows2005}, is a Wolter-I design grazing-incidence telescope, sensitive to X-rays in the range 0.3 - 10 keV.
Observations taken between 2016 December 1,  and 2017 April 15, (57723 - 57858 MJD) were taken as part of a multiwavelength monitoring campaign on \source.
In total, 55 observations were collected and analyzed.

\par \xrt\ observations were analyzed using the \texttt{HEAsoft}\footnote{\protect\url{https://heasarc.gsfc.nasa.gov/docs/software/lheasoft/}} (v6.31.1) software package. XRT event files were cleaned and calibrated using \texttt{xrtpipeline} (v0.13.7) \citep[see][]{capalbi2005swift}.
Since all of the observations within this period were taken in Windowed Timing mode (WT), 
the effects of pileup are considered to be negligible and no correction was applied \citep[see][]{Moretti2004,Romano2006}.
A circular source region centered on the location of OJ~287 was used, with radius 20 pixels ($\sim 47\arcsec$), encompassing 90\% of the point spread function (PSF) for a 1.5 keV photon.
An annular background region centered on the location of \source, with inner and outer radii of 80 and 120 pixels, respectively, was used.
Ancillary response files were generated using the \texttt{xrtmkarf} protocol and the \texttt{swxwt0to2s6\_20131212v015.rmf} response matrix was used.

\par Spectral analysis was performed using the \texttt{XSpec} package (12.9.1p) via the \texttt{PyXSpec} interface.
Data from each observation were grouped together, requiring at least 20 counts per energy bin to allow for $\chi^2$-fitting.
The energy spectra were fitted by an absorbed power-law model (\texttt{phabs * powerlaw}) assuming a neutral hydrogen density ($N_{HI}$) as measured by the LAB survey \citep{kalberla2005} of $2.49\times10^{20} \text{ cm$^{-2}$}$.
The photon index was observed to vary from $\Gamma \sim -(2.0 - 2.7)$.
This is softer than values typically observed for \source. 
This soft X-ray nature has been previously discussed by, for example, \citet{Kushwaha_2018, Huang_2021}.
This is in contrast to the observation by \citet{2018MNRAS.473.3638G} of a negative spectral curvature term ($\beta < 0$) in the hard X-ray band, indicating a hardening with increasing energy. This hardening with increasing energy was interpreted by \citet{2018MNRAS.473.3638G} as a transition from synchrotron to inverse-Compton dominance in the X-ray regime.

\par To test for spectral curvature, energy spectra were also fitted with an absorbed log-parabola model (\texttt{phabs * logpar}). 
Preference for a log-parabola model over a simple power-law model was tested using an F-test. 
Only 18 out of the 55 analyzed spectra showed a preference for a curved spectrum at the 95\% confidence level.
Of the 18 observations, 5 showed preference for a log parabola with $\beta > 0$ (positive/downwards curvature), with the remainder showing preference for a log parabola with $\beta <0$ (negative/upwards curvature).
Positive/downwards curvature ($\beta > 0$) occurs on the nights of 57783, 57786, 57788, 57792 and 57799 MJD. These occur just before and after the X-ray peak.
The presence of a softening with energy suggests that synchrotron emission is the dominant emission process in the X-ray regime during this flaring period.

The absorption-corrected integral flux was obtained in three energy bands approximately equi-spaced in log-space using the \texttt{cflux} protocol. 
These energy bands are labeled Soft (0.3-1 keV), Moderate (1-3 keV) and Hard (3-10 keV). 
These are shown in Panel (c) of Figure~\ref{fig:MWL_LC}.

\subsubsection{\uvot}\label{sec:swift:uvot}

The \uvot\ observations of OJ 287 between MJD 57719 and 57846 were analyzed using standard \texttt{HEAsoft} (v6.12) tools.\footnote{Data were processed using the pipeline available at \url{https://github.com/KarlenS/swift-uvot-analysis-tools}}
The Level 2 calibrated UVOT data, including sky images, were obtained from the High Energy Astrophysics Science Archive Research Center (HEASARC)\footnote{\url{https://heasarc.gsfc.nasa.gov}} data archives. UVOT observations for OJ 287 are available in the six UVOT bands: V, B, U, UVW1, UVW2, UVM2.
Aperture photometry was performed on these images using the \texttt{uvotsource} protocol. A circular aperture or radius 5$''$ centered on \source~was used for the source region, and a 20$''$ circular aperture centered on a source-free patch of the sky near the source region was used as a background region. 
The aperture photometry was used to produce a light curve for each available band, as shown in Figure~\ref{fig:MWL_LC}. For deriving the SED points, the central wavelength values from \citet{poole2008} were used for each band. The UVOT fluxes were corrected for Galactic extinction using dust reddening estimates from \citet{schlafly2011}.

\subsection{Optical}\label{obs:opt}
Optical observations were collected by numerous telescopes: 
(1) the 1.54 m Kuiper and 2.3 m Bok telescopes of Steward Observatory (Mt. Bigelow and Kitt Peak, AZ), hereinafter ``Steward''; 
(2) the 1.83 m Perkins telescope of Lowell Observatory (Flagstaff, AZ), hereinafter ``Perkins''; 
(3) the Robotically Controlled Telescope (RCT) located at Kitt Peak National Observatory (Tucson, AZ), hereinafter ``RCT'';
(4) the T-11 at New Mexico Skies Observatories (Mayhill NM), hereinafter ``T-11''.

Data were taken in the Johnson B, V, R, and I bands by the Steward and Perkins telescopes. They were processed and reduced using the methods described in \cite{Jorstad_2010,Jorstad_2013}.
Johnson R-band observations taken by the RCT were processed and reduced using the methods described in  \cite{Strolger_2014}.
Johnson R-band observations taken by the T-11 were taken as part of the iTelescope network.\footnote{\url{http://www.itelescope.net}}

The optical light curves are shown in Figure \ref{fig:optlight}, with the R-band polarization percentage and electron vector pointing angle (EVPA) shown in Figure \ref{fig:optpolar}.


\section{Temporal Analysis}\label{temporal}

\subsection{Multiwavelength Light Curves}\label{temporal:light}

\begin{figure}
  \plotone{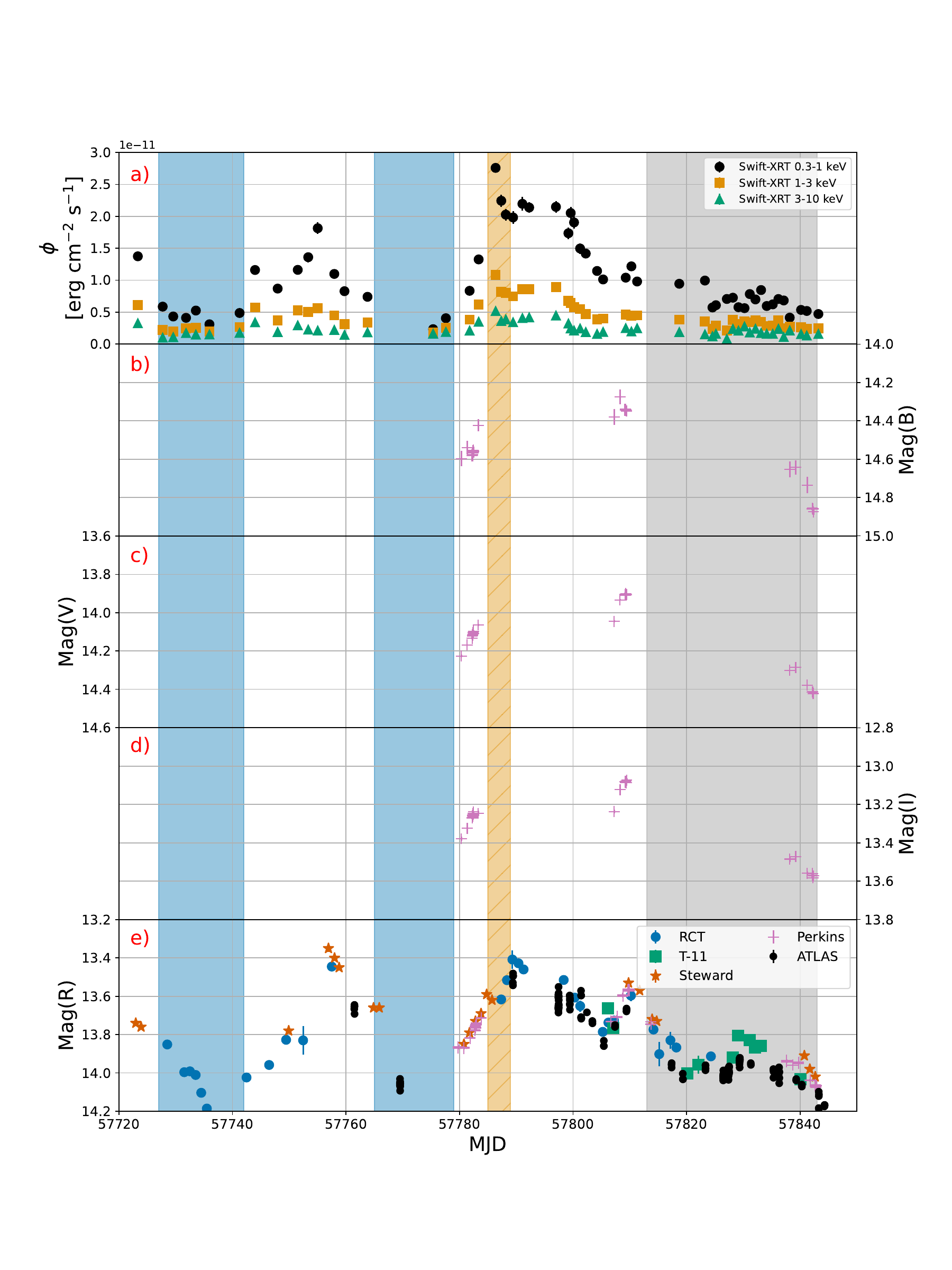}
  \caption{ X-ray and Optical light curve of \source. Panel A) shows the X-ray flux  as described in Figure \ref{fig:MWL_LC}; Panel b) shows the B-band magnitude; Panel c) shows the V-band magnitude; Panel d) shows the I-band magnitude and Panel e) shows the R-band magnitude. 
            \label{fig:optlight}
        }
\end{figure}

The multiwavelength light curves spanning UV-VHE and optical wavelengths are shown in Figures \ref{fig:MWL_LC} and \ref{fig:optlight}. 
A quantitative search for correlation and time-lags is discussed in Section \ref{cor:dcf}.
Here we first summarize some qualitative features of the various datasets. 
 A variability pattern with some similarities between X-ray and optical wavelengths is observed.
Visual inspection of the light curve shows a moderate increase is observed in both soft and hard X-rays at $\sim$57744 MJD, while there is no corresponding activity observed in any other band.
A steady rise from $\sim$57740 MJD, with a local maximum in the soft X-ray flux appears at $\sim$57755 MJD, before decreasing. 
After this decrease, fluxes rise with a global maximum in the soft X-ray occurring at $\sim$57785 MJD. 
The maximum soft X-ray flux corresponds to $\sim5\times$ the base level flux observed during this campaign and is temporally coincident with the VHE detection as reported by \citet{2017ATel10051....1M}.
Upon reaching this level at $\sim$57785 MJD, fluxes slowly decline towards the end of the observation period at $\sim$57850 MJD. 

\par In order to determine an upper bound on the minimum variability timescale, a scan is performed over the soft X-ray light curve. 
The minimum variability timescale is defined as the minimum time on which the flux doubles.
To prevent errors due to statistical fluctuations, a threshold of $3\sigma$ is applied to the significance of the flux change, with the significance estimated using Equation \ref{eqn:fluxChangeSigma}.
The minimum variability timescale is determined to be 2.7 days (57741.2 - 57744.0 MJD, 8.0 $\sigma$).

\subsection{Optical Polarization}\label{temporal:polar}

\begin{figure}
  \plotone{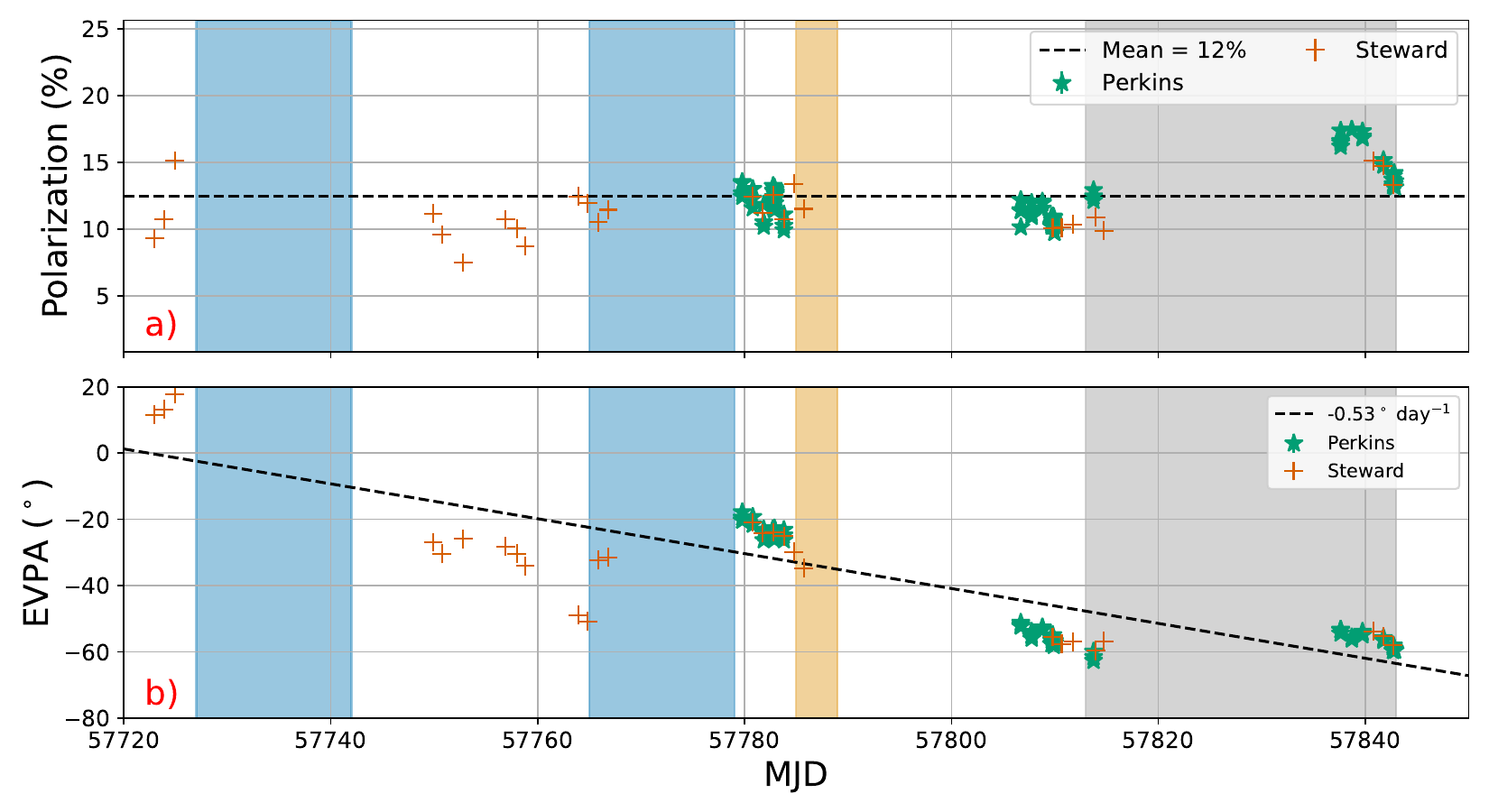}
  \caption{ Optical polarization measurements \source. The top panel shows the polarization percentage. The bottom panel shows the EVPA degree. See Section \ref{temporal:polar} text for details on annotations. 
            \label{fig:optpolar}
        }
\end{figure}

The mean degree of polarization is 12\% and represented in Figure \ref{fig:optpolar} Panel a) by the blacked dashed line.
The polarization remains approximately constant, varying by $\sim$3\% over the observed period.
The electric vector position angle (EVPA) shows a general clockwise rotation of $-0.53\pm0.03^{\circ} \mathrm{day}^{-1}$.
This was estimated by fitting a straight line to the observations, as illustrated by a dashed black line in Figure \ref{fig:optpolar} Panel b).
Such a trend corresponds to a rotational period of 1.86$\pm$0.06 years. 

\par A moderate increase in the polarization is observed toward the end of the post-flare state ($\sim$57840 MJD). However due to the sparse monitoring of the optical polarization, it is difficult to assess the correlation of this with features at any other wavelength.

\section{Correlation Analysis}\label{cor}

\subsection{Discrete Correlation Function Analysis}\label{cor:dcf}
The discrete correlation function \citep[DCF,][]{Edelson1988}, provides a method to study the correlation between two different time series without interpolating the data. 
The DCF of two datasets $\vec{a}$ and $\vec{b}$ is given by:
\begin{eqnarray}
\label{eqn:DCF}
	\mathrm{UDCF}_{ij} &= \frac{\left(a_i - \bar{a}\right)\left(b_j - \bar{b}\right)}{\sqrt{\left(\sigma_a^2 - e_a^2\right)\left(\sigma_b^2 - e_b^2\right)}},\\
    \mathrm{DCF}(\tau) &= \frac{1}{M} \sum \mathrm{UDCF}_{ij},
\end{eqnarray}

where $\tau$ is the time-lag, $\bar{f}$  and $\sigma_{f}$  are the mean and the standard deviation of dataset $\vec{f}$, $e_f$ is the statistical error associated with dataset $\vec{f}$ and $UDCF_{ij}$ is the discrete correlation between the time-lag pair $\Delta t_{ij} = t_j - t_i$. 
Finally, $DCF(\tau)$ is the mean discrete correlation for $M$ time-lag pairs such that $\tau - \Delta\tau/2 \leq \Delta t_{ij} < \tau + \Delta\tau/2$.

\begin{figure}
  \plotone{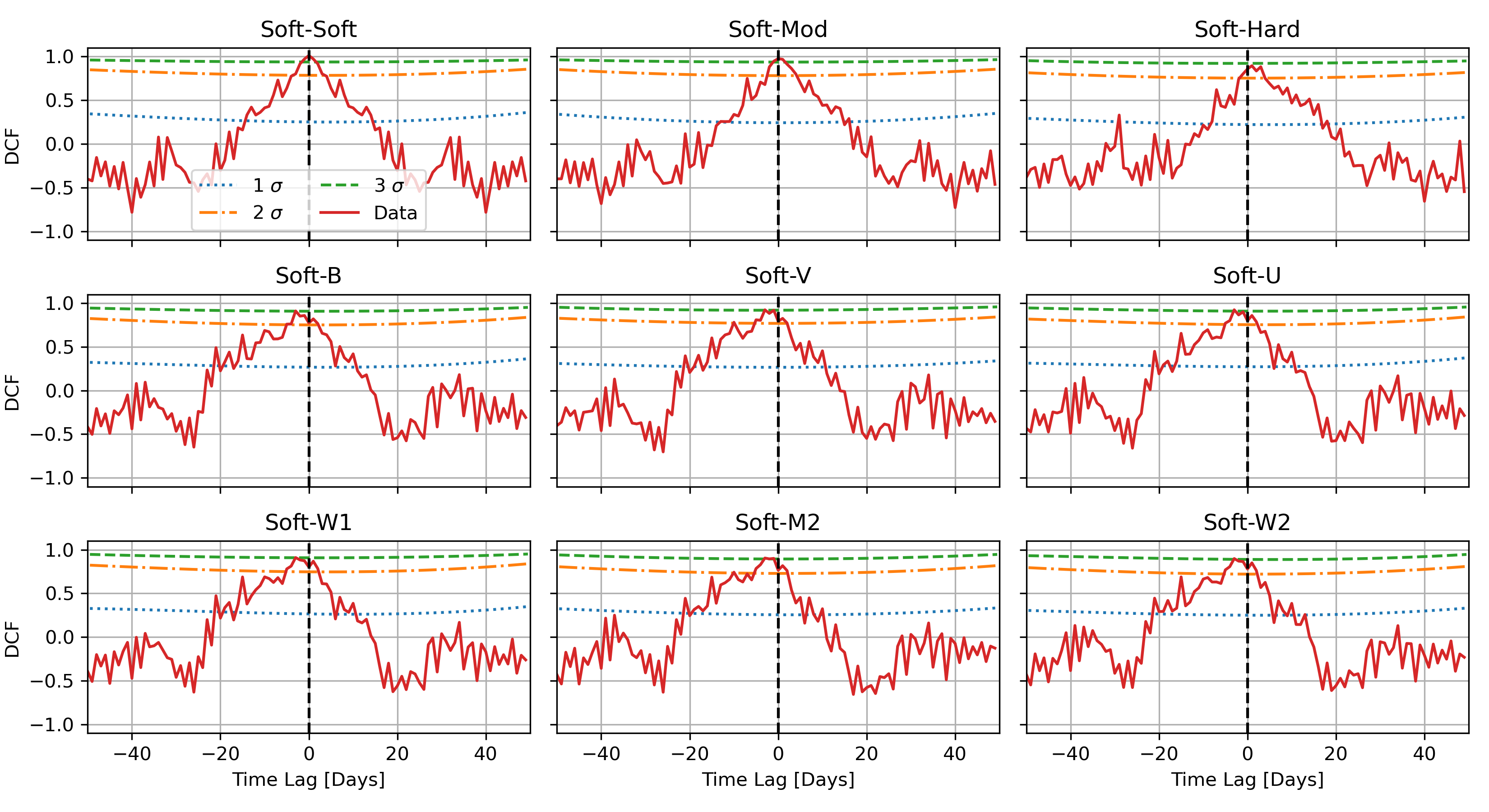}
   \caption{
  			DCF analysis for soft X-ray flux compared with other X-ray-UVOT flux bands. 
   			The top panel shows (left to right) Soft, Moderate and Hard X-ray flux.
   			The middle panel shows B, V and U.
            The bottom panel shows UVW1, UVM2 and UVW2.
   		    The dashed black line in each panel represents a time lag of zero days (no lag) between the two datasets.
            A negative time lag suggests a soft X-ray-led flare, while a positive one suggests a soft X-ray lag.
            The dotted blue, dot-dashed orange and dashed green lines represent the upper bound on the $1\sigma$, $2\sigma$- and $3\sigma$-confidence intervals.
            }
        \label{fig:SoftCorr}
\end{figure}

A DCF analysis is applied to the \uvot\ and \xrt\ light curves, with the results for the Soft X-ray flux band shown in Figure \ref{fig:SoftCorr}.
To determine the significance of an obtained DCF value, $10^5$ random light curves are generated for each flux band, using the methods described by \citet{Emmanoulopoulos2013} \citep[see also,][]{TimmerKoenig1995}.
The results were used to construct $1\sigma$-, $2\sigma$- and $3\sigma$-confidence intervals for the observed light curve pair.

\par Figure \ref{fig:SoftCorr} shows a correlation between the soft XRT band and the \uvot\ bands, significant at the $3\sigma$ level and a peak value occurring for a time lag of -1 to -3 days.
This suggests an X-ray-led time lag that is consistent with a visual inspection of Figure \ref{fig:MWL_LC}.
The broadness of the peak in the DCF is also consistent with zero time lag.  
The moderate X-ray band also shows a correlation with the UVOT bands at the $2-3\sigma$ level, while the hard X-ray flux only shows a weaker correlation.
The soft and moderate X-ray are correlated at zero time lag, however the hard X-ray flux shows weak $\sim2\sigma$ correlation with the soft X-ray.
The same weak correlation is also found between the hard X-ray and the moderate X-ray and \uvot\ fluxes.
This suggests that the dominant process for the soft and moderate X-ray emission is the same as the \uvot\ emission, while the hard X-ray emission may be due to a different process, or such that a correlation is below the sensitivity of the observations taken. While X-ray-led time-lags are suggested by the DCF, there is no clear difference in significance with the no-lag hypothesis. Hence we do not consider the time-lag evidence strong enough for further discussion.

\subsection{VHE-X-ray Correlation}\label{cor:vhe-xray}
Due to the low flux and Poissonian nature of the VHE observations, methods such as the DCF analysis presented in Section \ref{cor:dcf} are not suitable to compare the VHE and X-ray fluxes, as the errors cannot be assumed to be Gaussian.
Instead, a likelihood-based correlation analysis is applied.
This method assumes that a VHE-X-ray correlation can be described using a simple linear function such that:
\begin{equation}
\label{eqn:cor-vhe-xrt}
\phi_{VHE} = m\phi_{\text{X-ray}} + c,
\end{equation}
where $\phi_{i}$ is the flux in the $i$-th band, and $m$ and $c$ are parameters of a straight line fit.
For a set of VHE and X-ray observations, the likelihood of the VHE flux being described by Equation \ref{eqn:cor-vhe-xrt} can be obtained.
This is done by assuming the VHE spectral index is constant, therefore Equation \ref{eqn:cor-vhe-xrt} provides an estimate of the spectral normalization.
A joint-likelihood analysis is performed to find the maximum likelihood estimates $\hat{m}$ and $\hat{c}$.
The likelihood is then compared to the likelihood of a constant-flux model ($m = 0$), using the likelihood-ratio test.
This method allows for the inclusion of observations which resulted in upper limits, correctly calculating the Poisson likelihood for each observation.
In the case of a constant flux model, the likelihood-ratio test is expected to be $\chi^2$-distributed with one degree of freedom.
This was verified by Monte Carlo simulations of constant-flux and correlated-flux datasets simulated under similar observing conditions and flux levels to those observed during the observing campaign. 
This method does not consider uncertainties in the X-ray flux, which are much smaller fractionally than those for the VHE observations.

\par
\begin{figure}
  \plotone{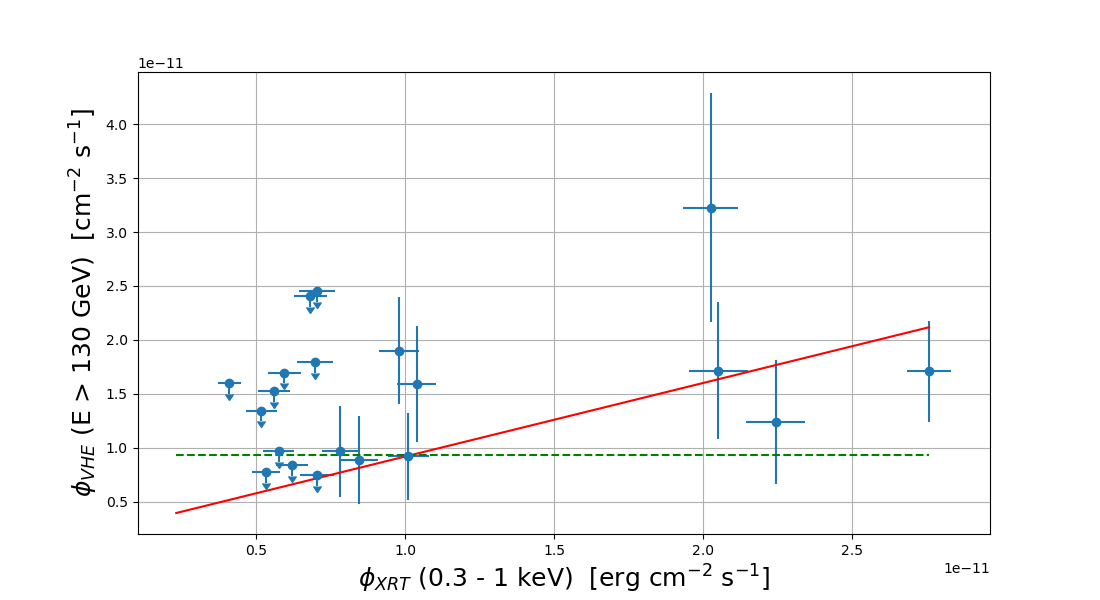}
  \caption{
  		Linear correlation between VHE and Soft (0.3-1 keV) X-ray flux.
        95\% C.L. upper limits are plotted for VHE observations with an excess significance $<2\sigma$, however they are included in the fit (see text).
        The red solid line corresponds to the best-fit linear correlation described by Equation \ref{eqn:cor-vhe-xrt}, with results shown in Table \ref{tab:vhe-xrt-cor} and the green dashed line shows the constant flux model.\label{fig:vhe-xrt-cor}
         }
\end{figure}

 The results of this correlation analysis are shown in Table \ref{tab:vhe-xrt-cor}.
The VHE flux shows a stronger correlation with the soft and moderate X-ray flux than with the hard X-ray flux. 
The VHE and soft/moderate X-ray flux shows a preference for a correlated-flux model  over a constant-flux model significant at the $>$99\% confidence level.
Figure \ref{fig:vhe-xrt-cor} shows the best-fit correlated- and constant-flux models for the VHE and soft X-ray data. The upper limits on the VHE flux are in disagreement with the constant-flux model. The inclusion of the upper-limit observations in the fit shows strong preference for the correlated-flux model.
This does not conflict with the constant-flux model in Section \ref{obs:ver}, as the details of the temporal order are included in the fit.

\startlongtable
\begin{deluxetable}{lccccc}
\tablecaption{Summary of the VHE-X-ray correlation results. Column 1 shows the corresponding X-ray dataset. Columns 2 and 3 show the best fit parameters assuming Equation \ref{eqn:cor-vhe-xrt}. Column 4 shows the corresponding log-likelihood value. Column 5 shows the likelihood ratio with respect to a constant flux model, with the equivalent $\chi^2$ probability shown in Column 6. \label{tab:vhe-xrt-cor}  }
\tablehead{
\colhead{XRT Data Set} & \colhead{Slope ($m$)} & \colhead{Constant ($c$)} & \colhead{$\log\mathcal{L}$} & \colhead{$2\log\left(\mathcal{L} / \mathcal{L}_0\right)$} & \colhead{Prob}  \\
\colhead{} & \colhead{$\left( [\mathrm{ergs^{-1}}] \right)$} & \colhead{$ \times 10 ^{-12} ~\left([\mathrm{cm^{-2} s^{-1}}]\right)$} & \colhead{} & \colhead{} & \colhead{}
}
\colnumbers
\startdata
    Constant Flux & $ 0 $ & $(9.35 \pm 1.19)$ &  $56066.49$\tablenotemark{a} & N/A & N/A \\
    Soft (0.3-1 keV) & $ (0.68 \pm 0.17) $ & $(2.36 \pm 2.61)$ & $ 56073.09 $ & $13.20$ & $0.0003$ \\
    Moderate (1-3 keV) & $ (1.95 \pm 0.35) $ & $(0.71 \pm 1.55)$ & $ 56072.77 $ & $ 12.56 $ & $0.0004$\\
    Hard (3-10 keV) & $ (4.04 \pm 0.54) $ & $(0.11 \pm 1.33)$ & $ 56071.10 $ & $9.22$ & $0.002$
    \enddata
\tablenotetext{a}{This corresponds to the null hypothesis of the likelihood-ratio test.}
\end{deluxetable}


\section{Broadband SED modeling}
\label{sed_modeling}
\subsection{Multi-wavelength states (Low, Flare, and Post-Flare)}

\begin{figure*}[ht!]
	 \begin{minipage}[b]{0.5\linewidth}
    \centering \includegraphics[width=9.4cm]{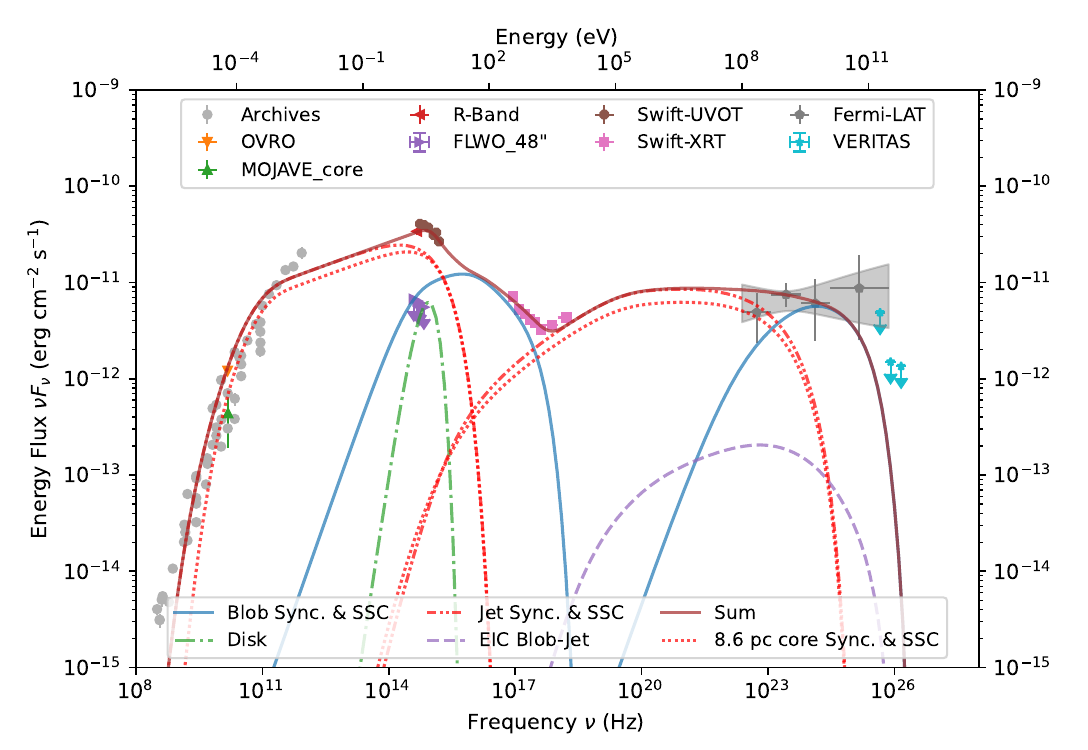}
    \put(-140,130){\makebox(0,0)[lb]{\textbf{Low}}}
	 \end{minipage}\hfill
	 \begin{minipage}[b]{0.5\linewidth}
   	\centering \includegraphics[width=9.4cm]{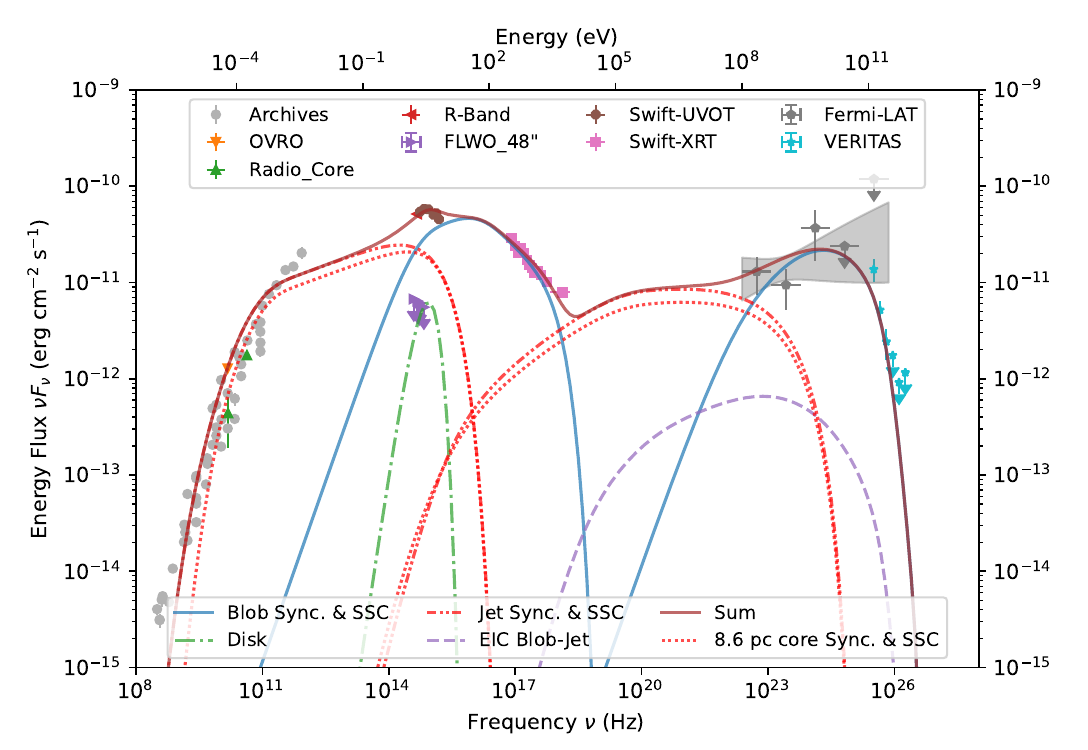}
   	\put(-140,130){\makebox(0,0)[lb]{\textbf{Flare}}}
  \end{minipage}\hfill
	\begin{minipage}[b]{0.5\linewidth}
    \centering \includegraphics[width=9.4cm]{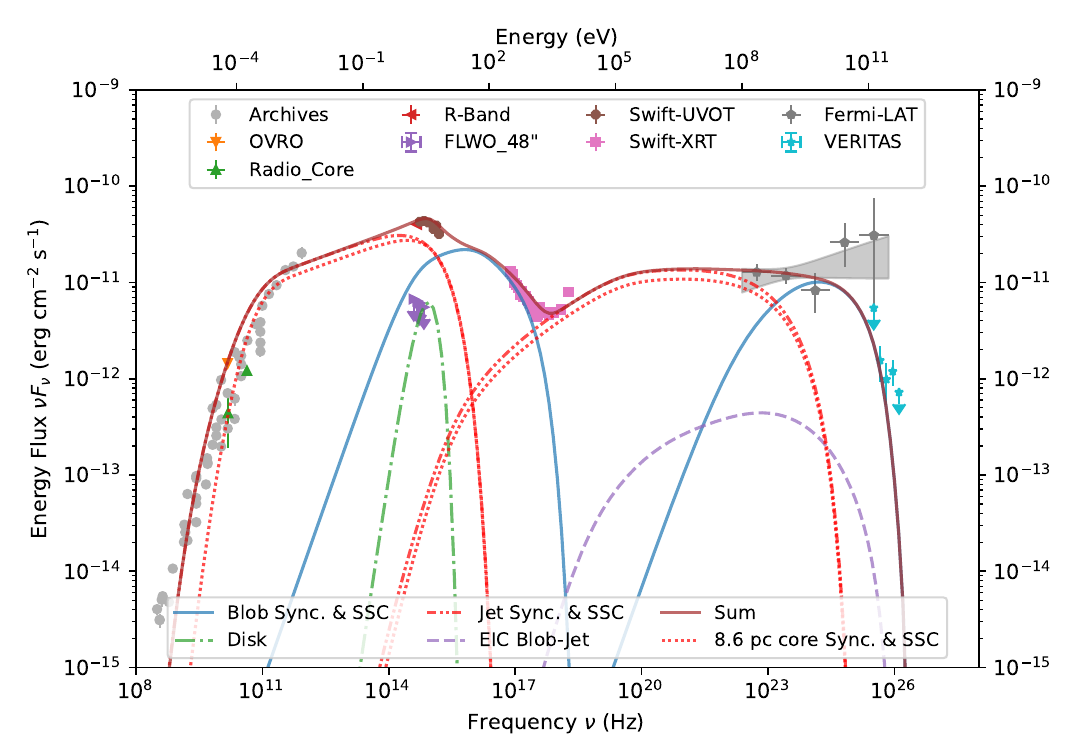}
    \put(-150,130){\makebox(0,0)[lb]{\textbf{Post-Flare}}}
  \end{minipage}\hfill
\caption{Broadband SEDs of OJ~287 Low, Flare, and Post-Flare activity states. The multi-zone radiative model \texttt{Bjet} is applied to the three activity states of the source. The $\gamma$-ray EBL absorption is taken into account considering the model of \cite{Franceschini_2017}. Archival radio data from various instruments and epochs are extracted with the ``SSDC SED Builder" \citep{Stratta_2011}. }
\label{fig:MWL_SED}
\end{figure*}

To better understand the multi-wavelength behavior of OJ~287, we build the broadband SED from radio to VHE for the three states of activity presented in Figure \ref{fig:MWL_LC}, namely Low (2016 December 5 - 20 and 2017 January 12 - 26), Flare (2017 February 1 - 5 ), and Post-Flare (2017 March 1 - 31).
Broadband SEDs are constructed for each of the three periods and are shown in Figure \ref{fig:MWL_SED}.
The Swift-XRT observations within each period are combined to produce a time-averaged spectrum.\footnote{See \url{https://swift.gsfc.nasa.gov/help/swiftfaq.html\#_xrt-combine}.} 
The VERITAS spectrum is obtained for each period by freezing the power-law index to the time-averaged fit and allowing the flux normalization to vary.

From these broadband SEDs, two main considerations can be made to guide the modeling approach.
First, no state of activity is Compton-dominated. The synchrotron peak reaches the maximum luminosity for all states. 
There are no hints that the observed flare is led by an external inverse-Compton process, which is usually associated with strongly Compton-dominated SEDs \cite[e.g.][]{Adams_2022}.
This is relatively unusual considering that most of the brightest flares of other TeV LBL/IBLs are at least moderately Compton-dominated, such as BL Lac \citep{Ravasio_2002,Abeysekara_2018}, W Comae \citep{Acciari_2009}, 3C 66A \citep{Abdo_2011}, or VER J0521+211 \citep{Adams_2022b}.

A second observation is that, despite soft X-rays that increase in flux by up to a factor of 10, the flare has a relatively marginal impact on other energy bands. We observe a steady flux in the R-band, 
and an increase of no more than a factor of 2 in optical-UV, hard X-rays ($> 3$ keV), HE, and VHE. 
Finally, 
 we observe a clear break in Low and Post-Flare X-ray spectra. Describing the observed soft X-ray variability and connecting the hard X-ray, Fermi, and VHE spectra in the SEDs in the frame of one-zone SSC appears unrealistic and led us to fully exclude this scenario in our study.

\subsection{Multi-zone approach}

In order to provide an interpretation of the intriguing MWL behavior of OJ 287, we use the multi-zone \texttt{Bjet} model \citep{Hervet2015, Hervet_2023}. We note that this model was initially used for the blazar AP Librae which shares similar features with OJ~287, such as being an LBL emitting in VHE with an extended X-ray jet and showing hybrid motion in its jet, with stationary and fast-moving knots observed in very-long-baseline interferometry (VLBI) radio observations.
This model considers a blob as a spherical compact high-energy zone moving through a larger, less dense, and slower section of a conical jet. It is based on the underlying assumption of a two-flow jet where the blob (fast spine perturbation) is radiatively interacting with the jet (steady slower sheath).
Both blob and jet are filled with a non-structured magnetic field and a non-thermal electron population. The electron spectrum of the blob has a broken power-law shape.
The jet is discretized in 50 cylindrical slices with a particle distribution set at the first slice as a simple power law. The particle density follows an adiabatic decrease for the downstream slices. The magnetic intensity, also set at the first slice, decreases with distance as $B(r) \propto r^{-1}$, as deduced from radio-VLBI core-shift measurements on a small sample of blazar jets by \cite{OSullivan_2009}. Both blob and jet emit synchrotron and SSC emissions. The radiation transfer of the blob emission through the jet is taken into account, as well as the inverse-Compton interaction of the blob particles on the surrounding jet synchrotron emission (``EIC blob-jet" in Figure \ref{fig:MWL_SED}).
The nucleus is also considered in \texttt{Bjet}, with an accretion disk simplified as a mono-temperature blackbody and a broad line region (BLR) reprocessing the disk radiation isotropically. We use the same BLR density profile and characterization as defined in \cite{Adams_2022}. The photon-photon opacity and the external inverse-Compton arising from the blob-BLR interaction are taken into account.

\begin{deluxetable}{ccccc}[ht!]
\tablecaption{Model parameters. Boldface values are the ones narrowed down to vary from one state to another.\label{tab::Params_model}}
\tabletypesize{\scriptsize}
    \tablehead{
    \colhead{Parameter} & \colhead{Low state} & \colhead{Flare} & \colhead{Post-Flare} & \colhead{Unit}\\ \hline
    \colhead{$\theta$} & \colhead{$2.0$} & $-$ & $-$ & \colhead{deg}\\ \hline
    \colhead{Blob} &  &
    }
    \startdata
    $\delta$ & $\mathbf{19}$ & $\mathbf{26.5}$ & $\mathbf{22}$\\ 
    $N_{e}^{(1)}$ & $1.9\times 10^{5}$ & $-$ & $-$ & cm$^{-3}$\\ 
    $n_1$ & $2.5$ & $-$ & $-$ & $-$ \\
    $n_2$ & $3.8$ & $-$ & $-$    & $-$\\
    $\gamma_{\mathrm{min}}$ & $4.0\times 10^{3}$ & $-$ & $-$ & $-$\\
    $\gamma_{\mathrm{max}}$ & $\mathbf{1.3\times 10^{5}}$ & $\mathbf{1.8\times 10^{5}}$ & $\mathbf{1.1\times 10^{5}}$ & $-$\\
    $\gamma_{\mathrm{brk}}$ & $2.3\times 10^{4}$ & $-$ & $-$ & $-$\\
    $B$ & $2.5\times 10^{-1}$ & $-$ & $-$ & G\\
    $R$ & $1.9\times 10^{16}$ & $-$ & $-$ & cm\\
    $D_{BH}$\tablenotemark{*} & $10$ & $-$ & $-$ & pc\\
    \hline
    \colhead{Nucleus} &  & \\
    \hline
    $L_{disk}$ & $2.5\times 10^{45}$ & $-$ & $-$ & erg s$^{-1}$\\ 
    $T_{disk}$ & $1.4\times 10^{4}$  & $-$ & $-$ & K\\
    \hline
    \colhead{Jet} &  & \\
    \hline
    $\delta$ & $15$ & $-$ & $-$\\ 
    $N_{e}^{(1)}$ & $\mathbf{1.3\times 10^{4}}$ & $-$ & $\mathbf{1.5\times 10^{4}}$  & cm$^{-3}$\\ 
    $n$ & $2.68$ & $-$ & $-$ & $-$\\
    $\gamma_{\mathrm{min}}$ & $1.0\times 10^{2}$ & $-$ & $-$ & $-$\\
    $\gamma_{\mathrm{max}}$ & $1.6\times 10^{4}$ & $-$ & $-$ & $-$\\
    $B_1$ & $2.5\times 10^{-1}$ & $-$ & $-$ & G\\
    $R_1$ & $6.9\times 10^{16}$ & $-$ & $-$ & cm\\
    $L$\tablenotemark{*} & $5.0\times 10^{1}$ & $-$ & $-$ & pc\\
    $\alpha/2$\tablenotemark{*} & $5.1\times 10^{-1}$ & $-$ & $-$ & deg
\enddata
\tablenotetext{}{$\theta$ is the angle of the blob direction of motion with respect to the
line of sight. The electron energy distribution between Lorentz factors $\gamma_{\mathrm{min}}$ and
$\gamma_{\mathrm{max}}$ is given by a broken power law with indices $-n_1$ and $-n_2$ below and above
$\gamma_{\mathrm{brk}}$ , with $N_{e}^{(1)}$ the normalization factor at $\gamma = 1$. The blob Doppler factor,
magnetic field, radius, and distance to the black hole are given by $\delta$, $B$, $R$, and
$D_{BH}$ , respectively. The disk luminosity and temperature are given by $L_{disk}$ and
$T_{disk}$. The jet is characterized by a length of $L$, and an opening angle of $\alpha$. Its radius and magnetic field strength are set for the first slice as $R_1$ and $B_1$, respectively.}
\tablenotetext{*}{\textit{ Host galaxy frame.}}
\label{tab::Bjet_parameters}
\end{deluxetable}

\subsection{Optical constraints on the accretion disk luminosity}

Accretion disks of AGN have long dynamic timescales. Considering the very conservative assumption of a dynamic scale given by the free-falling time from the outer edge of the accretion disk set at $1000 R_s$, we would get a minimum variability timescale for OJ~287 of
\begin{equation}
    \tau_{ff} = 1.27\times 10^{-7} \left( \frac{M_{BH}}{10 M_\odot} \right) = 228~ \rm{years},
\end{equation}
with the primary black hole mass of $M_{BH} = 1.8 \times 10^{10} M_\odot$.

Given the binary SMBH nature of OJ~287 we should consider the much shorter disk variability of $\sim 12$ years corresponding to the binary system periodicity, in the addition of disk flaring events happening during the crossing of the secondary black hole. The 2017 flare was right between two crossing events (2015.9 and 2019.6), and let us assume that the accretion disk was in a quiescent state at this time. Thanks to the deep monitoring of OJ~287 by the 48" optical telescope at the Fred Lawrence Whipple Observatory, with 156 observations taken between Dec. 2017 and Apr. 2024, we can pinpoint the lowest optical flux of the source during this period and use it as a constraint for the quiescent state of the accretion disk.
We find a minimum of optical activity in the Johnson-Cousin band (B,V) and the SDSS bands (r', i') on 2023, April 19. After galactic dust correction, we measure the lowest flux in the B band at $(5.34\pm 0.10) \times 10^{-12}$ erg cm$^{-2}$ s$^{-1}$. We consider the 2$\sigma$ confidence interval of all bands as flux upper limits for the accretion disk luminosity (see purple U.L. points in Figure \ref{fig:MWL_SED}). 
Our model use a simple mono-temperature blackbody emission to describe the accretion disk. From these constraints, we set up a disk luminosity of $2.5 \times 10^{45}$ erg s$^{-1}$ and a temperature of $1.4 \times 10^{4}$ Kelvin for all three states of activity.

\subsection{Parameters constrained with radio-VLBI observations}

We consider a jet angle with the line of sight of $\theta = 2.2\degree$, associated with a Doppler factor of $\delta = 23.9$, as estimated by \cite{Hervet2016} from the fastest observed apparent speed. These values are consistent with a different method from \cite{Jorstad2005}, where they estimate $\theta = 3.2\pm0.9$ and $\delta = 18.9\pm6.4$.

By considering the observed VLBI radio core as the base of the jet, we can constrain our jet model parameters geometrically and radiatively. 
From the public MOJAVE data \citep{Lister2019}, we derive an average FWHM of the radio core in the jet direction of $7.4\times 10^{-2}$ mas. 
At the distance of the source, this translates to a projected 0.33 pc. From an angle with a line of sight of 2.2\degree, we deduce a radio-core length of about 8.6 pc ($0.33/\sin(\theta) = 8.6$).
The flux of the 15.3 GHz core is set at its mean value over the 21 years of MOJAVE data. Given its large flux variability over the years, we set the flux error as the RMSD of the flux dispersion, which gives $F_{core,\mathrm{15GHz}} = (4.4 \pm 2.5) \times 10^{-13}$ erg cm$^{-2}$ s$^{-1}$.
The full length of our conical jet model is set at $L = 50$ pc. We constrain the jet parameters to match the extended radio emission and hard X-ray flux (dashed-dotted red lines in Figure \ref{fig:MWL_SED}). In order to model a consistent radio core, we adjust our jet model to also match the radio core flux when truncated at 8.6 pc from the SMBH (dotted red lines in Figure \ref{fig:MWL_SED}).

\subsection{Modeling consistency}

Given the numerous free parameters in our multi-zone model and the well-known degeneracy in the parameters of SSC-based radiative codes, we performed a ``fit-by-eye" approach. Hence, the modeling results do not exhaust the full range of parameters and possible scenarios that could adequately represent the data within our model framework.

After setting the observationally constrained parameters, our modeling approach consists of narrowing the observed MWL variability to the fewest of key physical parameters (See Table \ref{tab::Bjet_parameters}). This allows us to propose the simplest underlying physical process responsible for the observed flare.
Given the one-day flare timescale observed in X-rays, we consider that only a compact zone is responsible for the observed flare. 

As shown in Figure \ref{fig:MWL_SED}, a conical, core-dominated, 50 pc jet can match the observed radio emission down to the lowest frequencies. This jet also produces intense X-ray-to-gamma SSC radiation that can be seen as a flux baseline for the most energetic X-rays measured by \textit{Swift}-XRT and for the low energy gamma rays of \textit{Fermi}-LAT observed during the low state. This hard X-ray baseline is a natural way of explaining why most of the variability is observed in soft X-rays. 

As our model is not time-dependent, we ensure that all states can be treated as independent snapshots by checking that the deduced fastest variability and cooling times are shorter than the fastest observed variability of 2.7 day. 
Considering the usual variability formula $\tau_{\mathrm{min}} = R (1+z)/(c \delta)$, the fastest possible variability of the flaring blob component $\tau_{\mathrm{min}}$ is less than 13h for the three studied states. The jet component has a minimum variability timescale of about 56h, where its radius is minimal. It is regarded as mostly steady during and between the three studied periods, with only a marginal change of its particle density applied to the Post-Flare state to better adjust to the hard X-rays.  

We calculate the blob cooling time at the electron energy $\gamma_e = \gamma_{\mathrm{break}}$, which is associated with the peaks of the synchrotron and SSC emission. As the fastest variability is observed in X-rays, above the synchrotron peak, this cooling time estimation is a conservative approach.
The cooling time associated with the full radiative output (synchrotron and SSC) can be expressed in the Thomson regime as
\begin{equation}
T_{\mathrm{cool}} (\gamma_e) = \frac{3 m_e c}{4 \sigma_T \gamma_e  (U'_B + U'_\mathrm{syn})},
\end{equation}
with $m_e$ the electron mass, $\sigma_T$ the Thomson cross section, and  $U'_B$, $U'_\mathrm{syn}$ as the energy density in the blob frame of the magnetic field and synchrotron field, respectively \citep[e.g.][]{Inoue_1996}.
All calculated radiative cooling times are shorter than $\tau_{\mathrm{min}}$ and range from 4h to 6h. 

\section{Discussion}
\label{Section::Discussion}
\subsection{Location of the high-energy zone in the jet}
\label{Section::location_HE_zone}


The fact that none of the SEDs is Compton-dominated suggests that any external inverse-Compton process is relatively weak during the studied periods of activity. 
By parametrizing the BLR density profile from the accretion disk luminosity \cite[see eq. 8 in][]{Adams_2022}, one can propose a minimal distance of the blob from the SMBH where the BLR-blob interaction stays consistent with the observed $\gamma$-ray spectrum.

\begin{figure}[ht!]
   	\centering \includegraphics[width=9cm]{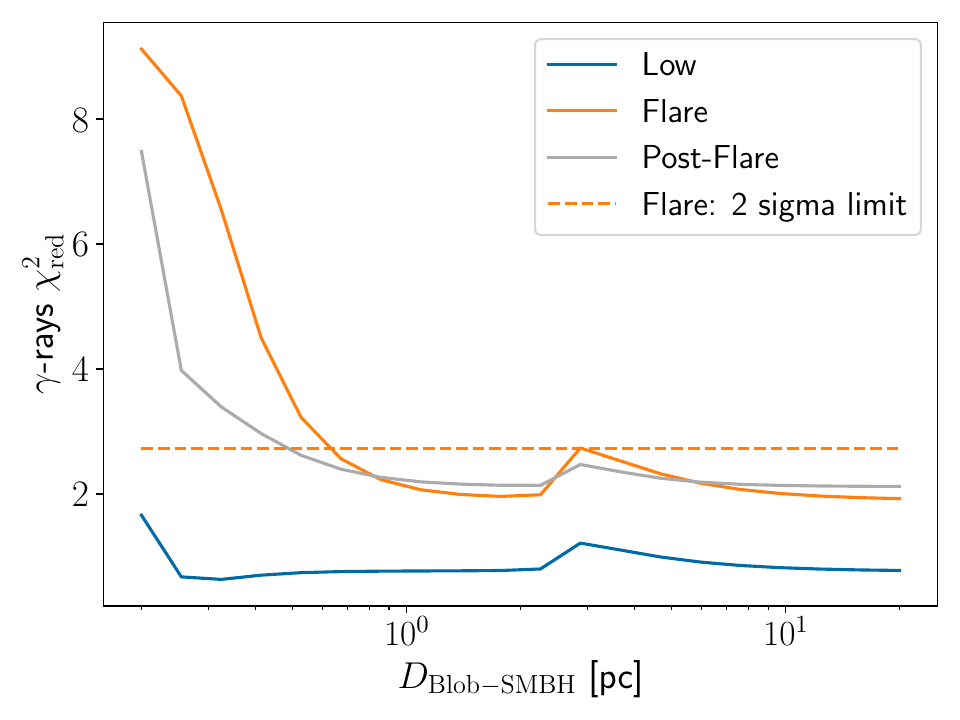}
\caption{Reduced $\chi^2$ of the $\gamma$-ray fitted SED for the three states of activity of OJ~287 according to the intrinsic distance of the blob from the SMBH ($D_{BH}$). The models are set to the parameters shown in Table \ref{tab::Params_model}, keeping free only the $D_{BH}$ value.   The domain above the orange dashed line represents the $\chi^2_\mathrm{red}$ values that are rejected at $>95.4 \%$ C.L (2 sigma) against the best $\chi^2_\mathrm{red}$ for the flaring state.  }
\label{fig:Chi2_blob_SMBH}
\end{figure}

To do so, we check the quality of the model representation in $\gamma$-rays for the three states at various blob-SMBH distances. We expect the fit to be poor at close distances where the photon-photon pair absorption prevents reaching the highest observed energies and when the EIC emission is significantly above the measured \textit{Fermi}-LAT flux.
We perform this check for the three states, considering only the distance as a free parameter and looking at the resulting reduced $\chi^2$ on the \textit{Fermi}-LAT+VERITAS dataset (see Figure \ref{fig:Chi2_blob_SMBH}). Without a measured VERITAS spectrum, the low state provides only weak constraints on the blob's location, compared to both flare and post-flare states. 
We see that at a short distance, the $\gamma$-ray emission suffers from strong photon-photon opacity that prevents reaching the observed VHE flux. As the blob gets closer to the brightest synchrotron part of the radio core ($\sim 3$ pc from the SMBH in our model), the EIC blob-jet emission gets brighter and increases the gamma-ray flux above the observed \textit{Fermi}-LAT spectrum which marginally impacts the fit quality.

From the Flare SED, our most constraining dataset, the location of the blob where the fitted SED is within a statistical level of $2 \sigma$ compared to the best fit is at a distance $D_{BH} \geq 0.6$ pc. A blob location within the densest part of the radio core is slightly disfavored, but our dataset cannot firmly reject this hypothesis.
We highlight that this result is heavily model-dependent and is entirely relevant only within the multi-zone model applied in this study.

\subsection{2017 flare: A shock in the first radio knot}

The complex flare of OJ 287 in February 2017 can be efficiently described by increasing the blob Doppler factor from 19 to 26.5 and the maximum electron energy $\gamma_\mathrm{max}$ from $1.3 \times 10^{5}$ to $1.8 \times 10^{5}$. The post-flare state is well modeled considering a Doppler factor decrease to 22 and $\gamma_\mathrm{max}$ decrease to $1.1 \times 10^{5}$. The hard X-ray increase between the Low and Post-Flare states can be modeled by a marginal particle density increase of the outer jet.
This behavior matches what we expect when an inner jet flow passes through a strong recollimation shock. Indeed, recollimation shocks are characterized by their upstream rarefaction waves that trigger a strong local flow acceleration. This effect can potentially increase the inner-jet bulk Lorentz factor to multiple times its nominal value \citep[e.g.][]{Mizuno_2015, Hervet_2017}. A blazar's plasmoid entering this zone should lead to a dramatic increase in its observed non-thermal emission.
The shocked plasmoid (or blob in our model) would undergo a high-energy particle injection and can display a higher maximum energy of accelerated particles, such as deduced from our modeling.

The lack of a significant external inverse-Compton signature in the studied MWL SEDs, as well as the requirement of a low gamma-gamma opacity from the VERITAS observations, lead us to favor such a shock downstream of the BLR at an unprojected distance of at least 0.6 pc from the SMBH (see Section \ref{Section::location_HE_zone}).
It has been noticed that most of the TeV IBLs and LBLs show radio VLBI stationary knots close to their radio core, which are often interpreted as stationary recollimation shocks \citep{Hervet2016}. OJ 287 is no exception, displaying stationary knots close to its core and fast-moving radio components downstream \citep{Lister2019}. This is consistent with a strong interplay between moving plasmoids and stationary shocks that has been observed multiple times to trigger large flares in similar sources, such as the eponymous BL Lacertae \citep{Marscher_2008, Abeysekara_2018}.

A recent study from \cite{Lico_2022} details observations performed by the 86 GHz Global Millimeter-VLBI Array (GMVA) around the time of the 2017 flare. They observed a brightening of the stationary radio knot S1 before the flare and the emergence of a new component K downstream of S1 after the flare.
Hence, they concluded it was very likely that the flare was triggered by a plasmoid crossing a recollimation shock defined by S1. With S1 being at the de-projected distance of $\sim 10$ pc from the core, their observations are fully consistent with our modeling results and interpretation.

\subsection{Contextualization with recent published studies of OJ 287}

As a blazar hosting an SMBH binary system and having large outbursts, OJ 287 has received considerable attention over the last decade. With numerous papers published every year, we do not intend to summarize the large number of multiwavelength studies performed on this source, but rather focus on the most recent and scientifically related works that will give a broader context to this paper.

The flare of 2017 has recently been studied by \cite{Huang_2021}, who suggested a possible origin as a tidal disruption event (TDE). The soft X-ray flare can indeed be fitted by a power law with a ``t$^{-5/3}$" decrease, as expected from TDEs. They also notice a relative strengthening of emission lines in the optical spectra of post-outburst epochs. This would be a strong argument for a nuclear origin. However, this relative strengthening seems to be caused rather by a dimming in the optical continuum than by a brightening of the lines. Given another soft X-ray flare in 2020 with similar properties \citep{Komossa_2021,Prince_2021}, it appears unlikely to have such a high rate of TDEs. The increase of VHE $\gamma$-ray flux observed by VERITAS  would also challenge a TDE origin, given the strong photon-photon absorption one would expect for such an event deep inside the BLR.

OJ 287 has also been modeled through methods applying a single SSC fit to a large number of blazars such as \cite{Lin_2018, Qin_2018}. Such an approach gives good statistical insight into understanding blazar populations. However, this approach usually fails to offer accurate descriptions of non-standard blazars such as OJ 287, as exemplified by the poor fit to the $\gamma$-ray SED in \cite{Qin_2018}.

\cite{Kushwaha_2018} performed modeling of the 2017 flare of OJ 287, making use of preliminary VERITAS results extracted from a plot in \cite{OBrien_2017}.  While they reach conclusions qualitatively similar to ours, we are able to be more quantitative with direct access to data and the final analysis.  For example, we are able to establish the correlation between X-rays and VHE emission at the $2 - 3\sigma$ level.

Working on \textit{Swift}-UVOT, \textit{Swift}-XRT, and \textit{Fermi}-LAT data, \cite{Prince_2021} calculated several single-zone SSC models for long periods ($> 100$ days) between 2017 and 2020. Their model and SED data show relatively good agreement. It is interesting to note that the spectral complexity of the source requiring a multi-zone model appears clearly only when having a broad multiwavelength coverage and focusing on specific activity states such as our dataset. This highlights the relevance of large, well-focused, multiwavelength campaigns.
Within this context, the ongoing, dedicated multiwavelength program MOMO (for Multiwavelength Observations and Modelling of OJ 287) from radio to X-rays is quite promising \citep{Komossa_2021b}. This is especially relevant to building a well-detailed MWL SED for other VHE outbursts similar to that observed by VERITAS.

\section{Conclusion}\label{con}

\label{Section::Conclusion}

After a long period of high X-ray activity beginning in 2016, the blazar OJ 287 went into an intense outburst phase that peaked during 2017 February 1-5. By monitoring this flare, VERITAS detected OJ 287 for the first time at VHE.
The timing of the flare does not appear related to any periodic activity triggered by the SMBH binary system or jet precession. Indeed, at the time of the flare, the secondary smaller black hole was behind the accretion disk of the primary one (from the observer's perspective), almost at the apogee between its previous and next disk crossing \cite[e.g.][]{Valtonen:2016dy}. Also, the beginning of 2017 corresponds to a period where the precession should make the jet the most misaligned with the line of sight \cite[][Figure 8]{Britzen2018, Huang_2021}.

A recent paper using VLBI radio data taken by the GMVA shows that the flare is consistent with a moving radio knot K crossing or ejected from a stationary bright radio zone S1 located about 10 pc (deprojected) from the core \cite{Lico_2022}.
All this information supports an intrinsic jet origin of the 2017 flare that is further confirmed by our study.

To investigate the 2017 outburst in-depth, we built a MWL lightcurve from December 2016 to March 2017, using data from \textit{Swift}, \textit{Fermi}, VERITAS, and multiple optical telescopes. We also retrieved optical polarization data from the Steward Observatory and the Lowell Observatory (Perkins).
Optical polarization did not highlight any specific behavior that would help us pinpoint the origin or location of the flare. The polarization percentage is relatively steady at $\sim 12\%$, with $3\%$ variations. The EVPA displays a continuous slow rotation through this period, corresponding to a rotational period of about two years. This period is much smaller than the SMBH orbital one ( $\sim 11$ years) or the jet precessing one ( $\sim 22$ years).

Broadband MWL light curves illustrate a complex MWL behavior, with a strong correlation observed between soft/moderate X-ray and optical-UV bands. All correlations are consistent with no time lags between wavelengths, although with a possible hint of X-ray-led flaring. 
The VHE emission is consistent with a constant-flux model on nightly time scales, but a significant ($>$99\% confidence) correlation is observed between the VHE and soft/moderate X-ray bands.

The most dramatic observation is the concentration of the flare radiation release in the soft X-ray band (0.3 - 1 KeV) with a factor of approximately 10 increase compared to the lowest observed flux within our period. In contrast, intermediate and hard X-rays only varied within a factor up to about 5 and 3, respectively.

This complexity led us to consider a leptonic multi-zone model to describe the broadband SED of OJ 287. To provide deeper insights into the flaring process, we built SEDs based on three selected states of activity: Low, Flare, and Post-Flare.
Our model considers the conical base of the jet (mainly observed as the radio-core) as a quasi-stationary component that emits in synchrotron and SSC, while the flaring component is pictured by a compact ``blob" moving through the jet, also emitting in synchrotron and SSC.

This model can satisfactorily match the MWL dataset for the three states of activity while being consistent with multiple observations such as the variability timescale, the radio core flux and extension, jet opening angle, and observed radio VLBI kinematics.
The complex variability observed can be reduced to two main varying parameters: the blob's Doppler factor and the maximum energy reached by the underlying particle distribution.
Our model favors very low photon-photon opacity to avoid suppressing VHE emission and provides a significantly better fit when the blob is situated outside the BLR ($>0.6$ pc from the SMBH) and marginally better when also outside of the radio core ($>5$ pc from the SMBH).

A strong increase of the Doppler factor outside of the radio core with an increase in particle maximum energies is an expected signature of a recollimation shock within the jet. The Doppler factor increase is a sign of a blob accelerating through the pre-shock rarefaction zone \citep[e.g.][]{Gomez_1997,Mizuno_2015,Hervet_2017}, while the particle energy increase is a sign of the post-shocked flow after a diffuse shock acceleration process \citep[e.g.][]{Meli_2013,Sironi_2015}. In 2020 another soft-X-ray flare was observed from \source~\citep[see, for example,][]{Komossa_2021, Prince_2021}, albeit dimmer than what was presented here. VERITAS was unable to observe \source~at the time of the flare, but this recent observation strengthened the idea of a typical behavior of OJ~287 rather than an exceptionally rare event.
Considering the radio VLBI observations of an interaction of the moving knot K and the stationary component S1 downstream of the radio core at the same time of the 2017 flare, we now have multiple pieces of evidence supporting recollimation shocks as major players in the high energy processes of OJ~287, as they may be for jetted AGN in general. 

Our study shows the tremendous scientific potential of ambitious MWL campaigns associated with radio-VLBI monitoring. Deepening such coordination with future ground-based observatories such as CTA, SKA-VLBI, and space telescopes will be of critical importance to further understand the jetted AGN phenomenon.

\vspace{5mm}
\facilities{VERITAS, Fermi(LAT), Swift(XRT and UVOT), }


\software{Event Display \citep{EDMaier2017},
		   VEGAS \citep{VEGASCogan2007},
          VEGAS-ITM \citep{christiansen2017},
          Science Tools \citep{2019ascl.soft05011F},
          Fermipy \citep{Wood:2017TJ},
          ROOT \citep{rene_brun_2019_3895860},
          Matplotlib \citep{Hunter:2007},
          NumPy \citep{harris2020array},
          SciPy \citep{2020SciPy-NMeth},
          AstroPy \citep{astropy:2013, astropy:2018, astropy:2022},
          Seaborn \citep{Waskom2021}
          }

\begin{acknowledgements}
  VERITAS is supported by grants from the U.S. Department of Energy Office of Science, the U.S. National Science Foundation and the Smithsonian Institution, and by NSERC in Canada.
  We acknowledge the excellent work of the technical support staff at the Fred Lawrence Whipple Observatory and at the collaborating institutions in the construction and operation of the instrument. O.H. thanks NSF for
support under grant PHY-2011420.
\end{acknowledgements}

\bibliography{OJ287_2018}

\begin{thebibliography}{}
\expandafter\ifx\csname natexlab\endcsname\relax\def\natexlab#1{#1}\fi

\bibitem[{{Abdo} {et~al.}(2011){Abdo}, {Ackermann}, {Ajello}, {Baldini},
  {Ballet}, {Barbiellini}, {Bastieri}, {Bechtol}, {Bellazzini}, {Berenji},
  {Blandford}, {Bonamente}, {Borgland}, {Bouvier}, {Bregeon}, {Brez},
  {Brigida}, {Bruel}, {Buehler}, {Buson}, {Caliandro}, {Cameron}, {Caraveo},
  {Carrigan}, {Casandjian}, {Cavazzuti}, {Cecchi}, {{\c{C}}elik}, {Charles},
  {Chekhtman}, {Cheung}, {Chiang}, {Ciprini}, {Claus}, {Cohen-Tanugi},
  {Conrad}, {Costamante}, {Cutini}, {Davis}, {Dermer}, {de Palma}, {Digel}, {do
  Couto e Silva}, {Drell}, {Dubois}, {Dumora}, {Favuzzi}, {Fegan}, {Fortin},
  {Frailis}, {Fuhrmann}, {Fukazawa}, {Funk}, {Fusco}, {Gargano}, {Gasparrini},
  {Gehrels}, {Germani}, {Giglietto}, {Giommi}, {Giordano}, {Giroletti},
  {Glanzman}, {Godfrey}, {Grenier}, {Grove}, {Guillemot}, {Guiriec}, {Hadasch},
  {Hayashida}, {Hays}, {Horan}, {Hughes}, {Itoh}, {J{\'o}hannesson}, {Johnson},
  {Johnson}, {Johnson}, {Kamae}, {Katagiri}, {Kataoka}, {Kn{\"o}dlseder},
  {Kuss}, {Lande}, {Latronico}, {Lee}, {Longo}, {Loparco}, {Lott},
  {Lovellette}, {Lubrano}, {Makeev}, {Mazziotta}, {McEnery}, {Mehault},
  {Michelson}, {Mizuno}, {Moiseev}, {Monte}, {Monzani}, {Morselli},
  {Moskalenko}, {Murgia}, {Nakamori}, {Naumann-Godo}, {Nestoras}, {Nolan},
  {Norris}, {Nuss}, {Ohsugi}, {Okumura}, {Omodei}, {Orlando}, {Ormes}, {Ozaki},
  {Paneque}, {Panetta}, {Parent}, {Pelassa}, {Pepe}, {Pesce-Rollins}, {Piron},
  {Porter}, {Rain{\`o}}, {Rando}, {Razzano}, {Reimer}, {Reimer}, {Reyes},
  {Ripken}, {Ritz}, {Romani}, {Roth}, {Sadrozinski}, {Sanchez}, {Sander},
  {Scargle}, {Sgr{\`o}}, {Shaw}, {Smith}, {Spandre}, {Spinelli}, {Strickman},
  {Suson}, {Takahashi}, {Tanaka}, {Thayer}, {Thayer}, {Thompson}, {Tibaldo},
  {Torres}, {Tosti}, {Tramacere}, {Usher}, {Vandenbroucke}, {Vasileiou},
  {Vilchez}, {Vitale}, {Waite}, {Wang}, {Winer}, {Wood}, {Yang}, {Ylinen},
  {Ziegler}, {Acciari}, {Aliu}, {Arlen}, {Aune}, {Beilicke}, {Benbow},
  {B{\"o}ttcher}, {Boltuch}, {Bradbury}, {Buckley}, {Bugaev}, {Byrum},
  {Cannon}, {Cesarini}, {Christiansen}, {Ciupik}, {Cui}, {de la Calle Perez},
  {Dickherber}, {Errando}, {Falcone}, {Finley}, {Finnegan}, {Fortson},
  {Furniss}, {Galante}, {Gall}, {Gillanders}, {Godambe}, {Grube}, {Guenette},
  {Gyuk}, {Hanna}, {Holder}, {Hui}, {Humensky}, {Imran}, {Kaaret}, {Karlsson},
  {Kertzman}, {Kieda}, {Konopelko}, {Krawczynski}, {Krennrich}, {Lang},
  {LeBohec}, {Maier}, {McArthur}, {McCann}, {McCutcheon}, {Moriarty},
  {Mukherjee}, {Ong}, {Otte}, {Pandel}, {Perkins}, {Pichel}, {Pohl}, {Quinn},
  {Ragan}, {Reynolds}, {Roache}, {Rose}, {Schroedter}, {Sembroski}, {Senturk},
  {Smith}, {Steele}, {Swordy}, {Te{\v{s}}i{\'c}}, {Theiling}, {Thibadeau},
  {Varlotta}, {Vassiliev}, {Vincent}, {Wakely}, {Ward}, {Weekes}, {Weinstein},
  {Weisgarber}, {Williams}, {Wissel}, {Wood}, {Villata}, {Raiteri}, {Gurwell},
  {Larionov}, {Kurtanidze}, {Aller}, {L{\"a}hteenm{\"a}ki}, {Chen},
  {Berduygin}, {Agudo}, {Aller}, {Arkharov}, {Bach}, {Bachev}, {Beltrame},
  {Ben{\'\i}tez}, {Buemi}, {Dashti}, {Calcidese}, {Capezzali}, {Carosati}, {Da
  Rio}, {Di Paola}, {Diltz}, {Dolci}, {Dultzin}, {Forn{\'e}}, {G{\'o}mez},
  {Hagen-Thorn}, {Halkola}, {Heidt}, {Hiriart}, {Hovatta}, {Hsiao}, {Jorstad},
  {Kimeridze}, {Konstantinova}, {Kopatskaya}, {Koptelova}, {Leto}, {Ligustri},
  {Lindfors}, {Lopez}, {Marscher}, {Mommert}, {Mujica}, {Nikolashvili},
  {Nilsson}, {Palma}, {Pasanen}, {Roca-Sogorb}, {Ros}, {Roustazadeh}, {Sadun},
  {Saino}, {Sigua}, {Sillan{\"a}{\"a}}, {Sorcia}, {Takalo}, {Tornikoski},
  {Trigilio}, {Turchetti}, {Umana}, {Belloni}, {Blake}, {Bloom}, {Angelakis},
  {Fumagalli}, {Hauser}, {Prochaska}, {Riquelme}, {Sievers}, {Starr},
  {Tagliaferri}, {Ungerechts}, {Wagner}, {Zensus}, {Fermi LAT Collaboration},
  {VERITAS Collaboration}, \& {GASP-WEBT Consortium}}]{Abdo_2011}
{Abdo}, A.~A., {Ackermann}, M., {Ajello}, M., {et~al.} 2011, \apj, 726, 43

\bibitem[{{Abdollahi} {et~al.}(2020){Abdollahi}, {Acero}, {Ackermann},
  {Ajello}, {Atwood}, {Axelsson}, {Baldini}, {Ballet}, {Barbiellini},
  {Bastieri}, {Becerra Gonzalez}, {Bellazzini}, {Berretta}, {Bissaldi},
  {Blandford}, {Bloom}, {Bonino}, {Bottacini}, {Brandt}, {Bregeon}, {Bruel},
  {Buehler}, {Burnett}, {Buson}, {Cameron}, {Caputo}, {Caraveo}, {Casandjian},
  {Castro}, {Cavazzuti}, {Charles}, {Chaty}, {Chen}, {Cheung}, {Chiaro},
  {Ciprini}, {Cohen-Tanugi}, {Cominsky}, {Coronado-Bl{\'a}zquez}, {Costantin},
  {Cuoco}, {Cutini}, {D'Ammando}, {DeKlotz}, {de la Torre Luque}, {de Palma},
  {Desai}, {Digel}, {Di Lalla}, {Di Mauro}, {Di Venere}, {Dom{\'\i}nguez},
  {Dumora}, {Fana Dirirsa}, {Fegan}, {Ferrara}, {Franckowiak}, {Fukazawa},
  {Funk}, {Fusco}, {Gargano}, {Gasparrini}, {Giglietto}, {Giommi}, {Giordano},
  {Giroletti}, {Glanzman}, {Green}, {Grenier}, {Griffin}, {Grondin}, {Grove},
  {Guiriec}, {Harding}, {Hayashi}, {Hays}, {Hewitt}, {Horan},
  {J{\'o}hannesson}, {Johnson}, {Kamae}, {Kerr}, {Kocevski}, {Kovac'evic'},
  {Kuss}, {Landriu}, {Larsson}, {Latronico}, {Lemoine-Goumard}, {Li},
  {Liodakis}, {Longo}, {Loparco}, {Lott}, {Lovellette}, {Lubrano}, {Madejski},
  {Maldera}, {Malyshev}, {Manfreda}, {Marchesini}, {Marcotulli},
  {Mart{\'\i}-Devesa}, {Martin}, {Massaro}, {Mazziotta}, {McEnery}, {Mereu},
  {Meyer}, {Michelson}, {Mirabal}, {Mizuno}, {Monzani}, {Morselli},
  {Moskalenko}, {Negro}, {Nuss}, {Ojha}, {Omodei}, {Orienti}, {Orlando},
  {Ormes}, {Palatiello}, {Paliya}, {Paneque}, {Pei}, {Pe{\~n}a-Herazo},
  {Perkins}, {Persic}, {Pesce-Rollins}, {Petrosian}, {Petrov}, {Piron}, {Poon},
  {Porter}, {Principe}, {Rain{\`o}}, {Rando}, {Razzano}, {Razzaque}, {Reimer},
  {Reimer}, {Remy}, {Reposeur}, {Romani}, {Saz Parkinson}, {Schinzel},
  {Serini}, {Sgr{\`o}}, {Siskind}, {Smith}, {Spandre}, {Spinelli}, {Strong},
  {Suson}, {Tajima}, {Takahashi}, {Tak}, {Thayer}, {Thompson}, {Tibaldo},
  {Torres}, {Torresi}, {Valverde}, {Van Klaveren}, {van Zyl}, {Wood},
  {Yassine}, \& {Zaharijas}}]{4FGL_2020}
{Abdollahi}, S., {Acero}, F., {Ackermann}, M., {et~al.} 2020, \apjs, 247, 33

\bibitem[{{Abeysekara} {et~al.}(2018){Abeysekara}, {Benbow}, {Bird},
  {Brantseg}, {Brose}, {Buchovecky}, {Buckley}, {Bugaev}, {Connolly}, {Cui},
  {Daniel}, {Falcone}, {Feng}, {Finley}, {Fortson}, {Furniss}, {Gillanders},
  {Gunawardhana}, {H{\"u}tten}, {Hanna}, {Hervet}, {Holder}, {Hughes},
  {Humensky}, {Johnson}, {Kaaret}, {Kar}, {Kertzman}, {Krennrich}, {Lang},
  {Lin}, {McArthur}, {Moriarty}, {Mukherjee}, {O'Brien}, {Ong}, {Otte}, {Park},
  {Petrashyk}, {Pohl}, {Pueschel}, {Quinn}, {Ragan}, {Reynolds}, {Richards},
  {Roache}, {Rulten}, {Sadeh}, {Santander}, {Sembroski}, {Shahinyan}, {Wakely},
  {Weinstein}, {Wells}, {Wilcox}, {Williams}, {Zitzer}, {VERITAS
  Collaboration}, {Jorstad}, {Marscher}, {Lister}, {Kovalev}, {Pushkarev},
  {Savolainen}, {Agudo}, {Molina}, {G{\'o}mez}, {Larionov}, {Borman},
  {Mokrushina}, {Tornikoski}, {L{\"a}hteenm{\"a}ki}, {Chamani}, {Enestam},
  {Kiehlmann}, {Hovatta}, {Smith}, \& {Pontrelli}}]{Abeysekara_2018}
{Abeysekara}, A.~U., {Benbow}, W., {Bird}, R., {et~al.} 2018, \apj, 856, 95

\bibitem[{Acciari {et~al.}(2008)Acciari, Beilicke, Blaylock, Bradbury, Buckley,
  Bugaev, Butt, Byrum, Celik, Cesarini, {et~al.}}]{acciari2008veritas}
Acciari, V., Beilicke, M., Blaylock, G., {et~al.} 2008, \apj, 679, 1427

\bibitem[{{Acciari} {et~al.}(2009){Acciari}, {Aliu}, {Aune}, {Beilicke},
  {Benbow}, {B{\"o}ttcher}, {Boltuch}, {Buckley}, {Bradbury}, {Bugaev},
  {Byrum}, {Cannon}, {Cesarini}, {Ciupik}, {Cogan}, {Cui}, {Dickherber},
  {Duke}, {Falcone}, {Finley}, {Fortin}, {Fortson}, {Furniss}, {Galante},
  {Gall}, {Gibbs}, {Gillanders}, {Grube}, {Guenette}, {Gyuk}, {Hanna},
  {Holder}, {Hui}, {Humensky}, {Kaaret}, {Karlsson}, {Kertzman}, {Kieda},
  {Konopelko}, {Krawczynski}, {Krennrich}, {Lang}, {Le Bohec}, {Maier},
  {McArthur}, {McCann}, {McCutcheon}, {Millis}, {Moriarty}, {Ong}, {Otte},
  {Pandel}, {Perkins}, {Pichel}, {Pohl}, {Quinn}, {Ragan}, {Reyes}, {Reynolds},
  {Roache}, {Rose}, {Sembroski}, {Smith}, {Steele}, {Theiling}, {Thibadeau},
  {Varlotta}, {Vassiliev}, {Vincent}, {Wakely}, {Ward}, {Weekes}, {Weinstein},
  {Weisgarber}, {Williams}, {Wissel}, {Wood}, {Pian}, {Vercellone},
  {Donnarumma}, {D'Ammando}, {Bulgarelli}, {Chen}, {Giuliani}, {Longo},
  {Pacciani}, {Pucella}, {Vittorini}, {Tavani}, {Argan}, {Barbiellini},
  {Caraveo}, {Cattaneo}, {Cocco}, {Costa}, {Del Monte}, {De Paris}, {Di Cocco},
  {Evangelista}, {Feroci}, {Fiorini}, {Froysland}, {Frutti}, {Fuschino},
  {Galli}, {Gianotti}, {Labanti}, {Lapshov}, {Lazzarotto}, {Lipari},
  {Marisaldi}, {Mastropietro}, {Mereghetti}, {Morelli}, {Morselli},
  {Pellizzoni}, {Perotti}, {Piano}, {Picozza}, {Pilia}, {Porrovecchio},
  {Prest}, {Rapisarda}, {Rappoldi}, {Rubini}, {Sabatini}, {Soffitta},
  {Trifoglio}, {Trois}, {Vallazza}, {Zambra}, {Zanello}, {Pittori},
  {Santolamazza}, {Verrecchia}, {Giommi}, {Colafrancesco}, {Salotti},
  {Villata}, {Raiteri}, {Aller}, {Aller}, {Arkharov}, {Efimova}, {Larionov},
  {Leto}, {Ligustri}, {Lindfors}, {Pasanen}, {Kurtanidze}, {Tetradze},
  {Lahteenmaki}, {Kotiranta}, {Cucchiara}, {Romano}, {Nesci}, {Pursimo},
  {Heidt}, {Benitez}, {Hiriart}, {Nilsson}, {Berdyugin}, {Mujica}, {Dultzin},
  {Lopez}, {Mommert}, {Sorcia}, \& {de la Calle Perez}}]{Acciari_2009}
{Acciari}, V.~A., {Aliu}, E., {Aune}, T., {et~al.} 2009, \apj, 707, 612

\bibitem[{{Adams} {et~al.}(2022{\natexlab{a}}){Adams}, {Batista}, {Benbow},
  {Brill}, {Brose}, {Buckley}, {Capasso}, {Christiansen}, {Errando}, {Feng},
  {Finley}, {Foote}, {Fortson}, {Furniss}, {Gallagher}, {Gent}, {Giuri},
  {Hanlon}, {Hanna}, {Hassan}, {Hervet}, {Holder}, {Hona}, {Hughes},
  {Humensky}, {Jin}, {Kaaret}, {Kertzman}, {Kieda}, {Kleiner}, {Krennrich},
  {Kumar}, {Lang}, {Lundy}, {Maier}, {Millis}, {Moriarty}, {Mukherjee},
  {Nievas-Rosillo}, {O'Brien}, {Ong}, {Otte}, {Patel}, {Patel}, {Pfrang},
  {Pohl}, {Prado}, {Pueschel}, {Quinn}, {Ragan}, {Reynolds}, {Ribeiro},
  {Roache}, {Ryan}, {Sadeh}, {Santander}, {Sembroski}, {Shang}, {Stevenson},
  {Tucci}, {Vassiliev}, {Wakely}, {Weinstein}, {Wells}, {Williams},
  {Williamson}, {Acciari}, {Aniello}, {Ansoldi}, {Antonelli}, {Arbet Engels},
  {Arcaro}, {Artero}, {Asano}, {Baack}, {Babi{\'c}}, {Baquero}, {Barres de
  Almeida}, {Barrio}, {Batkovi{\'c}}, {Becerra Gonz{\'a}lez}, {Bednarek},
  {Bernardini}, {Bernardos}, {Berti}, {Besenrieder}, {Bhattacharyya},
  {Bigongiari}, {Biland}, {Blanch}, {B{\"o}kenkamp}, {Bonnoli},
  {Bo{\v{s}}njak}, {Burelli}, {Busetto}, {Carosi}, {Ceribella}, {Cerruti},
  {Chai}, {Chilingarian}, {Cikota}, {Colombo}, {Contreras}, {Cortina},
  {Covino}, {D'Amico}, {D'Elia}, {Da Vela}, {Dazzi}, {De Angelis}, {De Lotto},
  {Del Popolo}, {Delfino}, {Delgado}, {Delgado Mendez}, {Depaoli}, {Di Pierro},
  {Di Venere}, {Do Souto Espi{\~n}eira}, {Dominis Prester}, {Donini}, {Dorner},
  {Doro}, {Elsaesser}, {Fallah Ramazani}, {Fari{\~n}a}, {Fattorini}, {Font},
  {Fruck}, {Fukami}, {Fukazawa}, {Garc{\'\i}a L{\'o}pez}, {Garczarczyk},
  {Gasparyan}, {Gaug}, {Giglietto}, {Giordano}, {Gliwny}, {Godinovi{\'c}},
  {Green}, {Green}, {Hadasch}, {Hahn}, {Hassan}, {Heckmann}, {Herrera},
  {Hrupec}, {H{\"u}tten}, {Inada}, {Iotov}, {Ishio}, {Iwamura}, {Jim{\'e}nez
  Mart{\'\i}nez}, {Jormanainen}, {Jouvin}, {Kerszberg}, {Kobayashi}, {Kubo},
  {Kushida}, {Lamastra}, {Lelas}, {Leone}, {Lindfors}, {Linhoff}, {Lombardi},
  {Longo}, {L{\'o}pez-Coto}, {L{\'o}pez-Moya}, {L{\'o}pez-Oramas}, {Loporchio},
  {Lorini}, {Machado de Oliveira Fraga}, {Maggio}, {Majumdar}, {Makariev},
  {Maneva}, {Manganaro}, {Mannheim}, {Mariotti}, {Mart{\'\i}nez}, {Mas
  Aguilar}, {Mazin}, {Menchiari}, {Mender}, {Mi{\'c}anovi{\'c}}, {Miceli},
  {Miener}, {Miranda}, {Mirzoyan}, {Molina}, {Mondal}, {Moralejo}, {Morcuende},
  {Moreno}, {Nakamori}, {Nanci}, {Nava}, {Neustroev}, {Nievas Rosillo},
  {Nigro}, {Nilsson}, {Nishijima}, {Noda}, {Nozaki}, {Ohtani}, {Oka},
  {Otero-Santos}, {Paiano}, {Palatiello}, {Paneque}, {Paoletti}, {Paredes},
  {Pavleti{\'c}}, {Pe{\~n}il}, {Persic}, {Pihet}, {Prada Moroni}, {Prandini},
  {Priyadarshi}, {Puljak}, {Rhode}, {Rib{\'o}}, {Rico}, {Righi}, {Rugliancich},
  {Sahakyan}, {Saito}, {Sakurai}, {Satalecka}, {Saturni}, {Schleicher},
  {Schmidt}, {Schmuckermaier}, {Schubert}, {Schweizer}, {Sitarek},
  {{\v{S}}nidari{\'c}}, {Sobczynska}, {Spolon}, {Stamerra},
  {Stri{\v{s}}kovi{\'c}}, {Strom}, {Strzys}, {Suda}, {Suri{\'c}}, {Takahashi},
  {Takeishi}, {Tavecchio}, {Temnikov}, {Terzi{\'c}}, {Teshima}, {Tosti},
  {Truzzi}, {Tutone}, {Ubach}, {Van Scherpenberg}, {Vanzo}, {Vazquez Acosta},
  {Ventura}, {Verguilov}, {Viale}, {Vigorito}, {Vitale}, {Vovk}, {Will},
  {Wunderlich}, {Yamamoto}, {Zari{\'c}}, \& {MAGIC
  Collaboration}}]{Adams_2022b}
{Adams}, C.~B., {Batista}, P., {Benbow}, W., {et~al.} 2022{\natexlab{a}}, \apj,
  932, 129

\bibitem[{{Adams} {et~al.}(2022{\natexlab{b}}){Adams}, {Benbow}, {Brill},
  {Buckley}, {Christiansen}, {Falcone}, {Feng}, {Finley}, {Foote}, {Fortson},
  {Furniss}, {Giuri}, {Hanna}, {Hassan}, {Hervet}, {Holder}, {Hona},
  {Humensky}, {Jin}, {Kaaret}, {Kleiner}, {Kumar}, {Lang}, {Lundy}, {Maier},
  {Moriarty}, {Mukherjee}, {Nievas Rosillo}, {O'Brien}, {Park}, {Patel},
  {Pfrang}, {Pohl}, {Prado}, {Pueschel}, {Quinn}, {Ragan}, {Reynolds},
  {Ribeiro}, {Roache}, {Ryan}, {Santander}, {Weinstein}, {Williams}, \&
  {Williamson}}]{2022A&A...658A..83A}
{Adams}, C.~B., {Benbow}, W., {Brill}, A., {et~al.} 2022{\natexlab{b}}, \aap,
  658, A83

\bibitem[{{Adams} {et~al.}(2022{\natexlab{c}}){Adams}, {Batshoun}, {Benbow},
  {Brill}, {Buckley}, {Capasso}, {Cavins}, {Christiansen}, {Coppi}, {Errando},
  {Farrell}, {Feng}, {Finley}, {Foote}, {Fortson}, {Furniss}, {Gent}, {Giuri},
  {Hanna}, {Hassan}, {Hervet}, {Holder}, {Houck}, {Humensky}, {Jin}, {Kaaret},
  {Kertzman}, {Kieda}, {Krennrich}, {Kumar}, {Lundy}, {Maier}, {McGrath},
  {Moriarty}, {Mukherjee}, {Nieto}, {Nievas-Rosillo}, {O'Brien}, {Ong},
  {Oppenheimer}, {Otte}, {Patel}, {Pfrang}, {Pohl}, {Prado}, {Pueschel},
  {Quinn}, {Ragan}, {Reynolds}, {Rhatigan}, {Ribeiro}, {Roache}, {Ryan},
  {Santander}, {Sembroski}, {Williams}, {Williamson}, {Valverde}, {Horan},
  {Buson}, {Cheung}, {Ciprini}, {Gasparrini}, {Ojha}, {van Zyl}, \&
  {Sironi}}]{Adams_2022}
{Adams}, C.~B., {Batshoun}, J., {Benbow}, W., {et~al.} 2022{\natexlab{c}},
  \apj, 924, 95

\bibitem[{{Archambault} {et~al.}(2021){Archambault}, {Chernitsky}, {Griffin},
  \& {Hanna}}]{2021APh...12802556A}
{Archambault}, S., {Chernitsky}, G., {Griffin}, S., \& {Hanna}, D. 2021,
  Astroparticle Physics, 128, 102556

\bibitem[{{Archambault} {et~al.}(2016){Archambault}, {Archer}, {Benbow},
  {Bird}, {Biteau}, {Buchovecky}, {Buckley}, {Bugaev}, {Byrum}, {Cerruti},
  {Chen}, {Ciupik}, {Connolly}, {Cui}, {Eisch}, {Errando}, {Falcone}, {Feng},
  {Finley}, {Fleischhack}, {Fortin}, {Fortson}, {Furniss}, {Gillanders},
  {Griffin}, {Grube}, {Gyuk}, {H{\"u}tten}, {H{\r{a}}kansson}, {Hanna},
  {Holder}, {Humensky}, {Johnson}, {Kaaret}, {Kar}, {Kelley-Hoskins},
  {Kertzman}, {Kieda}, {Krause}, {Krennrich}, {Kumar}, {Lang}, {Maier},
  {McArthur}, {McCann}, {Meagher}, {Moriarty}, {Mukherjee}, {Nguyen}, {Nieto},
  {O'Faol{\'a}in de Bhr{\'o}ithe}, {Ong}, {Otte}, {Park}, {Perkins}, {Pichel},
  {Pohl}, {Popkow}, {Pueschel}, {Quinn}, {Ragan}, {Reynolds}, {Richards},
  {Roache}, {Rovero}, {Santander}, {Sembroski}, {Shahinyan}, {Smith},
  {Staszak}, {Telezhinsky}, {Tucci}, {Tyler}, {Vincent}, {Wakely}, {Weiner},
  {Weinstein}, {Williams}, {Zitzer}, {VERITAS Collaboration}, {Fumagalli}, \&
  {Prochaska}}]{Archambault:2016ha}
{Archambault}, S., {Archer}, A., {Benbow}, W., {et~al.} 2016, \aj, 151, 142

\bibitem[{{Astropy Collaboration} {et~al.}(2013){Astropy Collaboration},
  {Robitaille}, {Tollerud}, {Greenfield}, {Droettboom}, {Bray}, {Aldcroft},
  {Davis}, {Ginsburg}, {Price-Whelan}, {Kerzendorf}, {Conley}, {Crighton},
  {Barbary}, {Muna}, {Ferguson}, {Grollier}, {Parikh}, {Nair}, {Unther},
  {Deil}, {Woillez}, {Conseil}, {Kramer}, {Turner}, {Singer}, {Fox}, {Weaver},
  {Zabalza}, {Edwards}, {Azalee Bostroem}, {Burke}, {Casey}, {Crawford},
  {Dencheva}, {Ely}, {Jenness}, {Labrie}, {Lim}, {Pierfederici}, {Pontzen},
  {Ptak}, {Refsdal}, {Servillat}, \& {Streicher}}]{astropy:2013}
{Astropy Collaboration}, {Robitaille}, T.~P., {Tollerud}, E.~J., {et~al.} 2013,
  \aap, 558, A33

\bibitem[{{Astropy Collaboration} {et~al.}(2018){Astropy Collaboration},
  {Price-Whelan}, {Sip{\H{o}}cz}, {G{\"u}nther}, {Lim}, {Crawford}, {Conseil},
  {Shupe}, {Craig}, {Dencheva}, {Ginsburg}, {Vand erPlas}, {Bradley},
  {P{\'e}rez-Su{\'a}rez}, {de Val-Borro}, {Aldcroft}, {Cruz}, {Robitaille},
  {Tollerud}, {Ardelean}, {Babej}, {Bach}, {Bachetti}, {Bakanov}, {Bamford},
  {Barentsen}, {Barmby}, {Baumbach}, {Berry}, {Biscani}, {Boquien}, {Bostroem},
  {Bouma}, {Brammer}, {Bray}, {Breytenbach}, {Buddelmeijer}, {Burke},
  {Calderone}, {Cano Rodr{\'\i}guez}, {Cara}, {Cardoso}, {Cheedella}, {Copin},
  {Corrales}, {Crichton}, {D'Avella}, {Deil}, {Depagne}, {Dietrich}, {Donath},
  {Droettboom}, {Earl}, {Erben}, {Fabbro}, {Ferreira}, {Finethy}, {Fox},
  {Garrison}, {Gibbons}, {Goldstein}, {Gommers}, {Greco}, {Greenfield},
  {Groener}, {Grollier}, {Hagen}, {Hirst}, {Homeier}, {Horton}, {Hosseinzadeh},
  {Hu}, {Hunkeler}, {Ivezi{\'c}}, {Jain}, {Jenness}, {Kanarek}, {Kendrew},
  {Kern}, {Kerzendorf}, {Khvalko}, {King}, {Kirkby}, {Kulkarni}, {Kumar},
  {Lee}, {Lenz}, {Littlefair}, {Ma}, {Macleod}, {Mastropietro}, {McCully},
  {Montagnac}, {Morris}, {Mueller}, {Mumford}, {Muna}, {Murphy}, {Nelson},
  {Nguyen}, {Ninan}, {N{\"o}the}, {Ogaz}, {Oh}, {Parejko}, {Parley}, {Pascual},
  {Patil}, {Patil}, {Plunkett}, {Prochaska}, {Rastogi}, {Reddy Janga},
  {Sabater}, {Sakurikar}, {Seifert}, {Sherbert}, {Sherwood-Taylor}, {Shih},
  {Sick}, {Silbiger}, {Singanamalla}, {Singer}, {Sladen}, {Sooley},
  {Sornarajah}, {Streicher}, {Teuben}, {Thomas}, {Tremblay}, {Turner},
  {Terr{\'o}n}, {van Kerkwijk}, {de la Vega}, {Watkins}, {Weaver}, {Whitmore},
  {Woillez}, {Zabalza}, \& {Astropy Contributors}}]{astropy:2018}
{Astropy Collaboration}, {Price-Whelan}, A.~M., {Sip{\H{o}}cz}, B.~M., {et~al.}
  2018, \aj, 156, 123

\bibitem[{{Astropy Collaboration} {et~al.}(2022){Astropy Collaboration},
  {Price-Whelan}, {Lim}, {Earl}, {Starkman}, {Bradley}, {Shupe}, {Patil},
  {Corrales}, {Brasseur}, {N{"o}the}, {Donath}, {Tollerud}, {Morris},
  {Ginsburg}, {Vaher}, {Weaver}, {Tocknell}, {Jamieson}, {van Kerkwijk},
  {Robitaille}, {Merry}, {Bachetti}, {G{"u}nther}, {Aldcroft},
  {Alvarado-Montes}, {Archibald}, {B{'o}di}, {Bapat}, {Barentsen}, {Baz{'a}n},
  {Biswas}, {Boquien}, {Burke}, {Cara}, {Cara}, {Conroy}, {Conseil}, {Craig},
  {Cross}, {Cruz}, {D'Eugenio}, {Dencheva}, {Devillepoix}, {Dietrich},
  {Eigenbrot}, {Erben}, {Ferreira}, {Foreman-Mackey}, {Fox}, {Freij}, {Garg},
  {Geda}, {Glattly}, {Gondhalekar}, {Gordon}, {Grant}, {Greenfield}, {Groener},
  {Guest}, {Gurovich}, {Handberg}, {Hart}, {Hatfield-Dodds}, {Homeier},
  {Hosseinzadeh}, {Jenness}, {Jones}, {Joseph}, {Kalmbach}, {Karamehmetoglu},
  {Ka{l}uszy{'n}ski}, {Kelley}, {Kern}, {Kerzendorf}, {Koch}, {Kulumani},
  {Lee}, {Ly}, {Ma}, {MacBride}, {Maljaars}, {Muna}, {Murphy}, {Norman},
  {O'Steen}, {Oman}, {Pacifici}, {Pascual}, {Pascual-Granado}, {Patil},
  {Perren}, {Pickering}, {Rastogi}, {Roulston}, {Ryan}, {Rykoff}, {Sabater},
  {Sakurikar}, {Salgado}, {Sanghi}, {Saunders}, {Savchenko}, {Schwardt},
  {Seifert-Eckert}, {Shih}, {Jain}, {Shukla}, {Sick}, {Simpson},
  {Singanamalla}, {Singer}, {Singhal}, {Sinha}, {Sip{H{o}}cz}, {Spitler},
  {Stansby}, {Streicher}, {{{S}}umak}, {Swinbank}, {Taranu}, {Tewary},
  {Tremblay}, {Val-Borro}, {Van Kooten}, {Vasovi{'c}}, {Verma}, {de Miranda
  Cardoso}, {Williams}, {Wilson}, {Winkel}, {Wood-Vasey}, {Xue}, {Yoachim},
  {Zhang}, {Zonca}, \& {Astropy Project Contributors}}]{astropy:2022}
{Astropy Collaboration}, {Price-Whelan}, A.~M., {Lim}, P.~L., {et~al.} 2022,
  apj, 935, 167

\bibitem[{{Atwood} {et~al.}(2013){Atwood}, {Albert}, {Baldini}, {Tinivella},
  {Bregeon}, {Pesce-Rollins}, {Sgr{\`o}}, {Bruel}, {Charles}, {Drlica-Wagner},
  {Franckowiak}, {Jogler}, {Rochester}, {Usher}, {Wood}, {Cohen-Tanugi}, \&
  {S.~Zimmer for the Fermi-LAT Collaboration}}]{Atwood2013}
{Atwood}, W., {Albert}, A., {Baldini}, L., {et~al.} 2013, ArXiv e-prints,
  arXiv:1303.3514

\bibitem[{{Atwood} {et~al.}(2009){Atwood}, {Abdo}, {Ackermann}, {Althouse},
  {Anderson}, {Axelsson}, {Baldini}, {Ballet}, {Band}, {Barbiellini}, \&
  et~al.}]{Atwood2009}
{Atwood}, W.~B., {Abdo}, A.~A., {Ackermann}, M., {et~al.} 2009, \apj, 697, 1071

\bibitem[{{Boettcher}(2012)}]{2012arXiv1205.0539B}
{Boettcher}, M. 2012, arXiv e-prints, arXiv:1205.0539

\bibitem[{{Britzen} {et~al.}(2018){Britzen}, {Fendt}, {Witzel}, {Qian},
  {Pashchenko}, {Kurtanidze}, {Zajacek}, {Martinez}, {Karas}, {Aller}, {Aller},
  {Eckart}, {Nilsson}, {Ar{\'e}valo}, {Cuadra}, {Subroweit}, \&
  {Witzel}}]{Britzen2018}
{Britzen}, S., {Fendt}, C., {Witzel}, G., {et~al.} 2018, \mnras, 478, 3199

\bibitem[{Brun {et~al.}(2019)Brun, Rademakers, Canal, Naumann, Couet, Moneta,
  Vassilev, Linev, Piparo, GANIS, Bellenot, Guiraud, Amadio, wverkerke, Mato,
  TimurP, Tadel, wlav, Tejedor, Blomer, Gheata, Hageboeck, Roiser, marsupial,
  Wunsch, Shadura, Bose, CristinaCristescu, Valls, \&
  Isemann}]{rene_brun_2019_3895860}
Brun, R., Rademakers, F., Canal, P., {et~al.} 2019, root-project/root:
  v6.18/02, doi:10.5281/zenodo.3895860

\bibitem[{{Burrows} {et~al.}(2005){Burrows}, {Hill}, {Nousek}, {Kennea},
  {Wells}, {Osborne}, {Abbey}, {Beardmore}, {Mukerjee}, {Short}, {Chincarini},
  {Campana}, {Citterio}, {Moretti}, {Pagani}, {Tagliaferri}, {Giommi},
  {Capalbi}, {Tamburelli}, {Angelini}, {Cusumano}, {Br{\"a}uninger}, {Burkert},
  \& {Hartner}}]{Burrows2005}
{Burrows}, D.~N., {Hill}, J.~E., {Nousek}, J.~A., {et~al.} 2005, \ssr, 120, 165

\bibitem[{Capalbi {et~al.}(2005)Capalbi, Perri, Saija, Tamburelli, \&
  Angelini}]{capalbi2005swift}
Capalbi, M., Perri, M., Saija, B., Tamburelli, F., \& Angelini, L. 2005, hosted
  by NASA at \url{https://swift.gsfc.nasa.gov/analysis/xrt\_swguide\_v1\_2.pdf}

\bibitem[{{Cerruti}(2020)}]{2020Galax...8...72C}
{Cerruti}, M. 2020, Galaxies, 8, 72

\bibitem[{{Cerruti} {et~al.}(2015){Cerruti}, {Zech}, {Boisson}, \&
  {Inoue}}]{2015MNRAS.448..910C}
{Cerruti}, M., {Zech}, A., {Boisson}, C., \& {Inoue}, S. 2015, \mnras, 448, 910

\bibitem[{Christiansen(2017)}]{christiansen2017}
Christiansen, J. 2017, in Proceedings of {{Science}} ({Trieste, Italy}: {Sissa
  Medialab}), 789

\bibitem[{{Cogan}(2007)}]{VEGASCogan2007}
{Cogan}, P. 2007, International Cosmic Ray Conference, 3, 1385

\bibitem[{{Dermer} \& {Schlickeiser}(1993)}]{1993ApJ...416..458D}
{Dermer}, C.~D., \& {Schlickeiser}, R. 1993, \apj, 416, 458

\bibitem[{{Edelson} {et~al.}(2015){Edelson}, {McHardy}, {Jorstad}, {Marscher},
  {Hovatta}, \& {Vaughan}}]{2015ATel.7056....1E}
{Edelson}, R., {McHardy}, I., {Jorstad}, S., {et~al.} 2015, The Astronomer's
  Telegram, 7056

\bibitem[{Edelson \& Krolik(1988)}]{Edelson1988}
Edelson, R.~A., \& Krolik, J.~H. 1988, \apj, 333, 646

\bibitem[{Emmanoulopoulos {et~al.}(2013)Emmanoulopoulos, McHardy, \&
  Papadakis}]{Emmanoulopoulos2013}
Emmanoulopoulos, D., McHardy, I.~M., \& Papadakis, I.~E. 2013, \mnras, 433, 907

\bibitem[{{Fermi Science Support Development Team}(2019)}]{2019ascl.soft05011F}
{Fermi Science Support Development Team}. 2019, {Fermitools: Fermi Science
  Tools}, Astrophysics Source Code Library, record ascl:1905.011, ascl:1905.011

\bibitem[{{Franceschini} \& {Rodighiero}(2017)}]{Franceschini_2017}
{Franceschini}, A., \& {Rodighiero}, G. 2017, \aap, 603, A34

\bibitem[{{Gaur} {et~al.}(2018){Gaur}, {Mohan}, {Wierzcholska}, \&
  {Gu}}]{2018MNRAS.473.3638G}
{Gaur}, H., {Mohan}, P., {Wierzcholska}, A., \& {Gu}, M. 2018, \mnras, 473,
  3638

\bibitem[{{Gehrels} {et~al.}(2004){Gehrels}, {Chincarini}, {Giommi}, {Mason},
  {Nousek}, {Wells}, {White}, {Barthelmy}, {Burrows}, {Cominsky}, {Hurley},
  {Marshall}, {M{\'e}sz{\'a}ros}, {Roming}, {Angelini}, {Barbier}, {Belloni},
  {Campana}, {Caraveo}, {Chester}, {Citterio}, {Cline}, {Cropper}, {Cummings},
  {Dean}, {Feigelson}, {Fenimore}, {Frail}, {Fruchter}, {Garmire}, {Gendreau},
  {Ghisellini}, {Greiner}, {Hill}, {Hunsberger}, {Krimm}, {Kulkarni}, {Kumar},
  {Lebrun}, {Lloyd-Ronning}, {Markwardt}, {Mattson}, {Mushotzky}, {Norris},
  {Osborne}, {Paczynski}, {Palmer}, {Park}, {Parsons}, {Paul}, {Rees},
  {Reynolds}, {Rhoads}, {Sasseen}, {Schaefer}, {Short}, {Smale}, {Smith},
  {Stella}, {Tagliaferri}, {Takahashi}, {Tashiro}, {Townsley}, {Tueller},
  {Turner}, {Vietri}, {Voges}, {Ward}, {Willingale}, {Zerbi}, \&
  {Zhang}}]{2004ApJ...611.1005G}
{Gehrels}, N., {Chincarini}, G., {Giommi}, P., {et~al.} 2004, \apj, 611, 1005

\bibitem[{{Ghisellini} {et~al.}(1996){Ghisellini}, {Maraschi}, \&
  {Dondi}}]{1996A&AS..120C.503G}
{Ghisellini}, G., {Maraschi}, L., \& {Dondi}, L. 1996, \aaps, 120, 503

\bibitem[{{G{\'o}mez} {et~al.}(1997){G{\'o}mez}, {Mart{\'\i}}, {Marscher},
  {Ib{\'a}{\~n}ez}, \& {Alberdi}}]{Gomez_1997}
{G{\'o}mez}, J.~L., {Mart{\'\i}}, J.~M., {Marscher}, A.~P., {Ib{\'a}{\~n}ez},
  J.~M., \& {Alberdi}, A. 1997, \apjl, 482, L33

\bibitem[{{Grupe} {et~al.}(2016){Grupe}, {Komossa}, \&
  {Gomez}}]{2016ATel.9629....1G}
{Grupe}, D., {Komossa}, S., \& {Gomez}, J.~L. 2016, The Astronomer's Telegram,
  9629

\bibitem[{{Hanna} {et~al.}(2022){Hanna}, {O'Brien}, \&
  {Rosin}}]{2022NIMPA102766235H}
{Hanna}, D., {O'Brien}, S., \& {Rosin}, T. 2022, Nuclear Instruments and
  Methods in Physics Research A, 1027, 166235

\bibitem[{Harris {et~al.}(2020)Harris, Millman, van~der Walt, Gommers,
  Virtanen, Cournapeau, Wieser, Taylor, Berg, Smith, Kern, Picus, Hoyer, van
  Kerkwijk, Brett, Haldane, del R{\'{i}}o, Wiebe, Peterson,
  G{\'{e}}rard-Marchant, Sheppard, Reddy, Weckesser, Abbasi, Gohlke, \&
  Oliphant}]{harris2020array}
Harris, C.~R., Millman, K.~J., van~der Walt, S.~J., {et~al.} 2020, Nature, 585,
  357

\bibitem[{{Hervet} {et~al.}(2015){Hervet}, {Boisson}, \& {Sol}}]{Hervet2015}
{Hervet}, O., {Boisson}, C., \& {Sol}, H. 2015, \aap, 578, A69

\bibitem[{{Hervet} {et~al.}(2016){Hervet}, {Boisson}, \& {Sol}}]{Hervet2016}
---. 2016, \aap, 592, A22

\bibitem[{{Hervet} {et~al.}(2023){Hervet}, {Johnson}, \&
  {Youngquist}}]{Hervet_2023}
{Hervet}, O., {Johnson}, C.~A., \& {Youngquist}, A. 2023, arXiv e-prints,
  arXiv:2307.08804

\bibitem[{{Hervet} {et~al.}(2017){Hervet}, {Meliani}, {Zech}, {Boisson},
  {Cayatte}, {Sauty}, \& {Sol}}]{Hervet_2017}
{Hervet}, O., {Meliani}, Z., {Zech}, A., {et~al.} 2017, \aap, 606, A103

\bibitem[{{Hillas} {et~al.}(1998){Hillas}, {Akerlof}, {Biller}, {Buckley},
  {Carter-Lewis}, {Catanese}, {Cawley}, {Fegan}, {Finley}, {Gaidos},
  {Krennrich}, {Lamb}, {Lang}, {Mohanty}, {Punch}, {Reynolds}, {Rodgers},
  {Rose}, {Rovero}, {Schubnell}, {Sembroski}, {Vacanti}, {Weekes}, {West}, \&
  {Zweerink}}]{1998ApJ...503..744H}
{Hillas}, A.~M., {Akerlof}, C.~W., {Biller}, S.~D., {et~al.} 1998, \apj, 503,
  744

\bibitem[{Holder {et~al.}(2006)Holder, Atkins, Badran, Blaylock, Bradbury,
  Buckley, Byrum, Carter-Lewis, Celik, Chow, {et~al.}}]{holder2006first}
Holder, J., Atkins, R., Badran, H., {et~al.} 2006, Astroparticle Physics, 25,
  391

\bibitem[{{Huang} {et~al.}(2021){Huang}, {Hu}, {Yin}, {Chen}, {Alexeeva},
  {Gao}, \& {Jiang}}]{Huang_2021}
{Huang}, S., {Hu}, S., {Yin}, H., {et~al.} 2021, \apj, 920, 12

\bibitem[{Hudec {et~al.}(2013)Hudec, Basta, Pihajoki, \& Valtonen}]{Hudec2013}
Hudec, R., Basta, M., Pihajoki, P., \& Valtonen, M. 2013, \aap, 559, 1

\bibitem[{Hunter(2007)}]{Hunter:2007}
Hunter, J.~D. 2007, Computing in Science \& Engineering, 9, 90

\bibitem[{{Inoue} \& {Takahara}(1996)}]{Inoue_1996}
{Inoue}, S., \& {Takahara}, F. 1996, \apj, 463, 555

\bibitem[{Johnston {et~al.}(1995)Johnston, Fey, Zacharias, Russell, Ma,
  de~Vegt, Reynolds, Jauncey, Archinal, Carter, Corbin, Eubanks, Florkowski,
  Hall, McCarthy, McCulloch, King, Nicolson, \& Shaffer}]{Johnston1995}
Johnston, K.~J., Fey, A.~L., Zacharias, N., {et~al.} 1995, \aj, 110, 880

\bibitem[{{Jorstad} {et~al.}(2005){Jorstad}, {Marscher}, {Lister}, {Stirling},
  {Cawthorne}, {Gear}, {G{\'o}mez}, {Stevens}, {Smith}, {Forster}, \&
  {Robson}}]{Jorstad2005}
{Jorstad}, S.~G., {Marscher}, A.~P., {Lister}, M.~L., {et~al.} 2005, \aj, 130,
  1418

\bibitem[{Jorstad {et~al.}(2010)Jorstad, Marscher, Larionov, Agudo, Smith,
  Gurwell, Lähteenmäki, Tornikoski, Markowitz, Arkharov, Blinov, Chatterjee,
  D'Arcangelo, Falcone, G{\'{o}}mez, Hagen-Thorn, Jordan, Kimeridze,
  Konstantinova, Kopatskaya, Kurtanidze, Larionova, Larionova, McHardy,
  Melnichuk, Roca-Sogorb, Schmidt, Skiff, Taylor, Thum, Troitsky, \&
  Wiesemeyer}]{Jorstad_2010}
Jorstad, S.~G., Marscher, A.~P., Larionov, V.~M., {et~al.} 2010, \apj, 715, 362

\bibitem[{Jorstad {et~al.}(2013)Jorstad, Marscher, Smith, Larionov, Agudo,
  Gurwell, Wehrle, Lähteenmäki, Nikolashvili, Schmidt, Arkharov, Blinov,
  Blumenthal, Casadio, Chigladze, Efimova, Eggen, G{\'{o}}mez, Grupe,
  Hagen-Thorn, Joshi, Kimeridze, Konstantinova, Kopatskaya, Kurtanidze,
  Kurtanidze, Larionova, Larionova, Sigua, MacDonald, Maune, McHardy, Miller,
  Molina, Morozova, Scott, Taylor, Tornikoski, Troitsky, Thum, Walker,
  Williamson, Sallum, Consiglio, \& Strelnitski}]{Jorstad_2013}
Jorstad, S.~G., Marscher, A.~P., Smith, P.~S., {et~al.} 2013, \apj, 773, 147

\bibitem[{{Kalberla} {et~al.}(2005){Kalberla}, {Burton}, {Hartmann}, {Arnal},
  {Bajaja}, {Morras}, \& {P{\"o}ppel}}]{kalberla2005}
{Kalberla}, P.~M.~W., {Burton}, W.~B., {Hartmann}, D., {et~al.} 2005, \aap,
  440, 775

\bibitem[{{Komossa} {et~al.}(2021{\natexlab{a}}){Komossa}, {Grupe}, {Gallo},
  {Gonzalez}, {Yao}, {Hollett}, {Parker}, \& {Ciprini}}]{Komossa_2021b}
{Komossa}, S., {Grupe}, D., {Gallo}, L.~C., {et~al.} 2021{\natexlab{a}}, \apj,
  923, 51

\bibitem[{{Komossa} {et~al.}(2021{\natexlab{b}}){Komossa}, {Grupe}, {Parker},
  {G{\'o}mez}, {Valtonen}, {Nowak}, {Jorstad}, {Haggard}, {Chandra}, {Ciprini},
  {Dey}, {Gopakumar}, {Hada}, {Markoff}, \& {Neilsen}}]{Komossa_2021}
{Komossa}, S., {Grupe}, D., {Parker}, M.~L., {et~al.} 2021{\natexlab{b}},
  \mnras, 504, 5575

\bibitem[{Krause {et~al.}(2017)Krause, Pueschel, \& Maier}]{Krause2017}
Krause, M., Pueschel, E., \& Maier, G. 2017, Astroparticle Physics, 89, 1

\bibitem[{{Kushwaha} {et~al.}(2018){Kushwaha}, {Gupta}, {Wiita}, {Pal}, {Gaur},
  {de Gouveia Dal Pino}, {Kurtanidze}, {Semkov}, {Damljanovic}, {Hu}, {Uemura},
  {Vince}, {Darriba}, {Gu}, {Bachev}, {Chen}, {Itoh}, {Kawabata}, {Kurtanidze},
  {Nakaoka}, {Nikolashvili}, {Sigua}, {Strigachev}, \& {Zhang}}]{Kushwaha_2018}
{Kushwaha}, P., {Gupta}, A.~C., {Wiita}, P.~J., {et~al.} 2018, \mnras, 479,
  1672

\bibitem[{{Laine} {et~al.}(2020){Laine}, {Dey}, {Valtonen}, {Gopakumar},
  {Zola}, {Komossa}, {Kidger}, {Pihajoki}, {G{\'o}mez}, {Caton}, {Ciprini},
  {Drozdz}, {Gazeas}, {Godunova}, {Haque}, {Hildebrandt}, {Hudec}, {Jermak},
  {Kong}, {Lehto}, {Liakos}, {Matsumoto}, {Mugrauer}, {Pursimo}, {Reichart},
  {Simon}, {Siwak}, \& {Sonbas}}]{2020ApJ...894L...1L}
{Laine}, S., {Dey}, L., {Valtonen}, M., {et~al.} 2020, \apjl, 894, L1

\bibitem[{Li \& Ma(1983)}]{Li1983}
Li, T.-P., \& Ma, Y.-Q. 1983, \apj, 272, 317

\bibitem[{{Lico} {et~al.}(2022){Lico}, {Casadio}, {Jorstad}, {G{\'o}mez},
  {Marscher}, {Traianou}, {Kim}, {Zhao}, {Fuentes}, {Cho}, {Krichbaum},
  {Hervet}, {O'Brien}, {Boccardi}, {Myserlis}, {Agudo}, {Alberdi}, {Weaver}, \&
  {Zensus}}]{Lico_2022}
{Lico}, R., {Casadio}, C., {Jorstad}, S.~G., {et~al.} 2022, \aap, 658, L10

\bibitem[{{Lin} \& {Fan}(2018)}]{Lin_2018}
{Lin}, C., \& {Fan}, J.-H. 2018, Research in Astronomy and Astrophysics, 18,
  120

\bibitem[{{Lister} {et~al.}(2019){Lister}, {Homan}, {Hovatta}, {Kellermann},
  {Kiehlmann}, {Kovalev}, {Max-Moerbeck}, {Pushkarev}, {Readhead}, {Ros}, \&
  {Savolainen}}]{Lister2019}
{Lister}, M.~L., {Homan}, D.~C., {Hovatta}, T., {et~al.} 2019, \apj, 874, 43

\bibitem[{{Lott} {et~al.}(2020){Lott}, {Gasparrini}, \&
  {Ciprini}}]{4LACDR2_2020}
{Lott}, B., {Gasparrini}, D., \& {Ciprini}, S. 2020, arXiv e-prints,
  arXiv:2010.08406

\bibitem[{{Maier} \& {Holder}(2017)}]{EDMaier2017}
{Maier}, G., \& {Holder}, J. 2017, International Cosmic Ray Conference,
  arXiv:1708.04048

\bibitem[{{Marscher} {et~al.}(2008){Marscher}, {Jorstad}, {D'Arcangelo},
  {Smith}, {Williams}, {Larionov}, {Oh}, {Olmstead}, {Aller}, {Aller},
  {McHardy}, {L{\"a}hteenm{\"a}ki}, {Tornikoski}, {Valtaoja}, {Hagen-Thorn},
  {Kopatskaya}, {Gear}, {Tosti}, {Kurtanidze}, {Nikolashvili}, {Sigua},
  {Miller}, \& {Ryle}}]{Marscher_2008}
{Marscher}, A.~P., {Jorstad}, S.~G., {D'Arcangelo}, F.~D., {et~al.} 2008, \nat,
  452, 966

\bibitem[{{Mattox} {et~al.}(1996){Mattox}, {Bertsch}, {Chiang}, {Dingus},
  {Digel}, {Esposito}, {Fierro}, {Hartman}, {Hunter}, {Kanbach}, {Kniffen},
  {Lin}, {Macomb}, {Mayer-Hasselwander}, {Michelson}, {von Montigny},
  {Mukherjee}, {Nolan}, {Ramanamurthy}, {Schneid}, {Sreekumar}, {Thompson}, \&
  {Willis}}]{1996ApJ...461..396M}
{Mattox}, J.~R., {Bertsch}, D.~L., {Chiang}, J., {et~al.} 1996, \apj, 461, 396

\bibitem[{{Meli} \& {Biermann}(2013)}]{Meli_2013}
{Meli}, A., \& {Biermann}, P.~L. 2013, \aap, 556, A88

\bibitem[{{Mizuno} {et~al.}(2015){Mizuno}, {G{\'o}mez}, {Nishikawa}, {Meli},
  {Hardee}, \& {Rezzolla}}]{Mizuno_2015}
{Mizuno}, Y., {G{\'o}mez}, J.~L., {Nishikawa}, K.-I., {et~al.} 2015, \apj, 809,
  38

\bibitem[{{Moretti} {et~al.}(2004){Moretti}, {Campana}, {Tagliaferri}, {Abbey},
  {Ambrosi}, {Angelini}, {Beardmore}, {Br{\"a}uninger}, {Burkert}, {Burrows},
  {Capalbi}, {Chincarini}, {Citterio}, {Cusumano}, {Freyberg}, {Giommi},
  {Hartner}, {Hill}, {Mori}, {Morris}, {Mukerjee}, {Nousek}, {Osborne},
  {Short}, {Tamburelli}, {Watson}, \& {Wells}}]{Moretti2004}
{Moretti}, A., {Campana}, S., {Tagliaferri}, G., {et~al.} 2004, in \procspie,
  Vol. 5165, X-Ray and Gamma-Ray Instrumentation for Astronomy XIII, ed. K.~A.
  {Flanagan} \& O.~H.~W. {Siegmund}, 232--240

\bibitem[{Mukherjee {et~al.}(2017)}]{2017ATel10051....1M}
Mukherjee, R., {et~al.} 2017, The Astronomer's Telegram, 10051

\bibitem[{{Nieppola, E.} {et~al.}(2006){Nieppola, E.}, {Tornikoski, M.}, \&
  {Valtaoja, E.}}]{Nieppola2006}
{Nieppola, E.}, {Tornikoski, M.}, \& {Valtaoja, E.} 2006, A\&A, 445, 441

\bibitem[{{Nilsson} {et~al.}(2010){Nilsson}, {Takalo}, {Lehto}, \&
  {Sillanp{\"a}{\"a}}}]{Nilsson2010}
{Nilsson}, K., {Takalo}, L.~O., {Lehto}, H.~J., \& {Sillanp{\"a}{\"a}}, A.
  2010, \aap, 516, A60

\bibitem[{{Nolan} {et~al.}(2012){Nolan}, {Abdo}, {Ackermann}, {Ajello},
  {Allafort}, {Antolini}, {Atwood}, {Axelsson}, {Baldini}, {Ballet}, \&
  et~al.}]{2FGL2012}
{Nolan}, P.~L., {Abdo}, A.~A., {Ackermann}, M., {et~al.} 2012, \apjs, 199, 31

\bibitem[{{O'Brien} {et~al.}(2017)}]{OBrien_2017}
{O'Brien}, S., {et~al.} 2017, in International Cosmic Ray Conference, Vol. 301,
  35th International Cosmic Ray Conference (ICRC2017), 650

\bibitem[{{O'Sullivan} \& {Gabuzda}(2009)}]{OSullivan_2009}
{O'Sullivan}, S.~P., \& {Gabuzda}, D.~C. 2009, \mnras, 400, 26

\bibitem[{{Padovani} \& {Giommi}(1995)}]{Padovani1995}
{Padovani}, P., \& {Giommi}, P. 1995, \apj, 444, 567

\bibitem[{{Park} {et~al.}(2015)}]{park2015performance}
{Park}, N., {et~al.} 2015, in International Cosmic Ray Conference, Vol.~34,
  34th International Cosmic Ray Conference (ICRC2015), 771

\bibitem[{Piron {et~al.}(2001)Piron, Djannati-Ata{\"{i}}, Punch, Tavernet,
  Barrau, Bazer-Bachi, Chounet, Debiais, Degrange, Dezalay, Espigat, Fabre,
  Fleury, Fontaine, Goret, Gouiffes, Khelifi, Malet, Masterson, Mohanty, Nuss,
  Renault, Rivoal, Rob, \& Vorobiov}]{Piron2001}
Piron, F., Djannati-Ata{\"{i}}, A., Punch, M., {et~al.} 2001, \aap, 374, 895

\bibitem[{{Poole} {et~al.}(2008){Poole}, {Breeveld}, {Page}, {Landsman},
  {Holland}, {Roming}, {Kuin}, {Brown}, {Gronwall}, {Hunsberger}, {Koch},
  {Mason}, {Schady}, {vanden Berk}, {Blustin}, {Boyd}, {Broos}, {Carter},
  {Chester}, {Cucchiara}, {Hancock}, {Huckle}, {Immler}, {Ivanushkina},
  {Kennedy}, {Marshall}, {Morgan}, {Pandey}, {de Pasquale}, {Smith}, \&
  {Still}}]{poole2008}
{Poole}, T.~S., {Breeveld}, A.~A., {Page}, M.~J., {et~al.} 2008, \mnras, 383,
  627

\bibitem[{{Prince} {et~al.}(2021){Prince}, {Agarwal}, {Gupta}, {Majumdar},
  {Czerny}, {Cellone}, \& {Andruchow}}]{Prince_2021}
{Prince}, R., {Agarwal}, A., {Gupta}, N., {et~al.} 2021, \aap, 654, A38

\bibitem[{{Qin} {et~al.}(2018){Qin}, {Wang}, {Yang}, {Yuan}, {Mao}, \&
  {Kang}}]{Qin_2018}
{Qin}, L., {Wang}, J., {Yang}, C., {et~al.} 2018, \pasj, 70, 5

\bibitem[{{Raiteri} {et~al.}(2010){Raiteri}, {Villata}, {Bruschini}, {Capetti},
  {Kurtanidze}, {Larionov}, {Romano}, {Vercellone}, {Agudo}, {Aller}, {Aller},
  {Arkharov}, {Bach}, {Berdyugin}, {Blinov}, {B{\"o}ttcher}, {Buemi},
  {Calcidese}, {Carosati}, {Casas}, {Chen}, {Coloma}, {Diltz}, {di Paola},
  {Dolci}, {Efimova}, {Forn{\'e}}, {G{\'o}mez}, {Gurwell}, {Hakola}, {Hovatta},
  {Hsiao}, {Jordan}, {Jorstad}, {Koptelova}, {Kurtanidze},
  {L{\"a}hteenm{\"a}ki}, {Larionova}, {Leto}, {Lindfors}, {Ligustri},
  {Marscher}, {Morozova}, {Nikolashvili}, {Nilsson}, {Ros}, {Roustazadeh},
  {Sadun}, {Sillanp{\"a}{\"a}}, {Sainio}, {Takalo}, {Tornikoski}, {Trigilio},
  {Troitsky}, \& {Umana}}]{2010A&A...524A..43R}
{Raiteri}, C.~M., {Villata}, M., {Bruschini}, L., {et~al.} 2010, \aap, 524, A43

\bibitem[{{Ravasio} {et~al.}(2002){Ravasio}, {Tagliaferri}, {Ghisellini},
  {Giommi}, {Nesci}, {Massaro}, {Chiappetti}, {Celotti}, {Costamante},
  {Maraschi}, {Tavecchio}, {Tosti}, {Treves}, {Wolter}, {Balonek}, {Carini},
  {Kato}, {Kurtanidze}, {Montagni}, {Nikolashvili}, {Noble}, {Nucciarelli},
  {Raiteri}, {Sclavi}, {Uemura}, \& {Villata}}]{Ravasio_2002}
{Ravasio}, M., {Tagliaferri}, G., {Ghisellini}, G., {et~al.} 2002, \aap, 383,
  763

\bibitem[{{Rolke} {et~al.}(2005){Rolke}, {L{\'o}pez}, \& {Conrad}}]{rolke2005}
{Rolke}, W.~A., {L{\'o}pez}, A.~M., \& {Conrad}, J. 2005, Nuclear Instruments
  and Methods in Physics Research A, 551, 493

\bibitem[{{Romano} {et~al.}(2006){Romano}, {Campana}, {Chincarini}, {Cummings},
  {Cusumano}, {Holland}, {Mangano}, {Mineo}, {Page}, {Pal'Shin}, {Rol},
  {Sakamoto}, {Zhang}, {Aptekar}, {Barbier}, {Barthelmy}, {Beardmore}, {Boyd},
  {Burrows}, {Capalbi}, {Fenimore}, {Frederiks}, {Gehrels}, {Giommi}, {Goad},
  {Godet}, {Golenetskii}, {Guetta}, {Kennea}, {La Parola}, {Malesani},
  {Marshall}, {Moretti}, {Nousek}, {O'Brien}, {Osborne}, {Perri}, \&
  {Tagliaferri}}]{Romano2006}
{Romano}, P., {Campana}, S., {Chincarini}, G., {et~al.} 2006, \aap, 456, 917

\bibitem[{{Schlafly} \& {Finkbeiner}(2011)}]{schlafly2011}
{Schlafly}, E.~F., \& {Finkbeiner}, D.~P. 2011, \apj, 737, 103

\bibitem[{Seta {et~al.}(2009)Seta, Isobe, Tashiro, Yaji, Arai, Fukuhara, Kohno,
  Nakanishi, Sasada, Shimajiri, Tosaki, Uemura, Anderhub, Antonelli, Antoranz,
  Backes, Baixeras, Balestra, Barrio, Bastieri, Gonz{\'{a}}lez, Becker,
  Bednarek, Berger, Bernardini, Biland, Bock, Bonnoli, Bordas, Tridon,
  Bosch-Ramon, Bose, Braun, Bretz, Britvitch, Camara, Carmona, Commichau,
  Contreras, Cortina, Dios, Covino, Curtef, Dazzi, DeAngelis, DelPozo,
  DelosReyes, DeLotto, DeMaria, DeSabata, M{\'{e}}ndez, Dom{\'{i}}nguez,
  Dorner, Doro, Elsaesser, Errando, Ferenc, Fern{\'{a}}ndez, Firpo, Fonseca,
  Font, Galante, Garc{\'{i}}aL{\'{o}}pez, Garczarczyk, Gaug, Goebel, Hadasch,
  Hayashida, Herrero, Hildebrand, H{\"{o}}hne-M{\"{o}}nch, Hose, Hsu, Jogler,
  Kranich, Barbera, Laille, Leonardo, Lindfors, Lombardi, Longo, L{\'{o}}pez,
  Lorenz, Majumdar, Maneva, Mankuzhiyil, Mannheim, Maraschi, Mariotti,
  Mart{\'{i}}nez, Mazin, Meucci, Meyer, Miranda, Mirzoyan, Miyamoto,
  Mold{\'{o}}n, Moles, Moralejo, Nieto, Nilsson, Ninkovic, Otte, Oya, Paoletti,
  Paredes, Pasanen, Pascoli, Pauss, Pegna, Perez-Torres, Persic, Peruzzo,
  Prada, Prandini, Puchades, Reichardt, Rhode, Rib{\'{o}}, Rico, Rissi, Robert,
  R{\"{u}}gamer, Saggion, Saito, Salvati, S{\'{a}}nchez-Conde, Satalecka,
  Scalzotto, Scapin, Schweizer, Shayduk, Shore, Sidro, Sierpowska-Bartosik,
  Sillanp{\"{a}}{\"{a}}, Sitarek, Sobczynska, Spanier, Stamerra, Schneebeli,
  Takalo, Tavecchio, Temnikov, Tescaro, Teshima, Tluczykont, Torres, Turini,
  Vankov, Wagner, Wittek, Zabalza, Zandanel, Zanin, \& Zapatero}]{Seta2009}
Seta, H., Isobe, N., Tashiro, M.~S., {et~al.} 2009, Publications of the
  Astronomical Society of Japan, 61, 1011

\bibitem[{{Shrader} {et~al.}(1996){Shrader}, {Hartman}, \&
  {Webb}}]{Egret_OJ287_1996A&AS}
{Shrader}, C.~R., {Hartman}, R.~C., \& {Webb}, J.~R. 1996, \aaps, 120, 599

\bibitem[{Sillanpaa {et~al.}(1988)Sillanpaa, Haarala, Valtonen, Sundelius, \&
  Byrd}]{Sillanpaa1988}
Sillanpaa, A., Haarala, S., Valtonen, M.~J., Sundelius, B., \& Byrd, G.~G.
  1988, \apj, 325, 628

\bibitem[{{Sironi} {et~al.}(2015){Sironi}, {Keshet}, \&
  {Lemoine}}]{Sironi_2015}
{Sironi}, L., {Keshet}, U., \& {Lemoine}, M. 2015, \ssr, 191, 519

\bibitem[{{Stratta} {et~al.}(2011){Stratta}, {Capalbi}, {Giommi}, {Primavera},
  {Cutini}, \& {Gasparrini}}]{Stratta_2011}
{Stratta}, G., {Capalbi}, M., {Giommi}, P., {et~al.} 2011, arXiv e-prints,
  arXiv:1103.0749

\bibitem[{Stroh \& Falcone(2013)}]{swift-mon2013}
Stroh, M.~C., \& Falcone, A.~D. 2013, \apjs, 207, 28

\bibitem[{Strolger {et~al.}(2014)Strolger, Gott, Carini, Engle, Gelderman,
  Guinan, Laney, McGruder, Treffers, \& Walter}]{Strolger_2014}
Strolger, L.-G., Gott, A.~M., Carini, M., {et~al.} 2014, \aj, 147, 49

\bibitem[{Sundelius {et~al.}(1997)Sundelius, Wahde, Lehto, \&
  Valtonen}]{Sundelius_1997}
Sundelius, B., Wahde, M., Lehto, H.~J., \& Valtonen, M.~J. 1997, \apj, 484, 180

\bibitem[{{Tavecchio} {et~al.}(1998){Tavecchio}, {Maraschi}, \&
  {Ghisellini}}]{1998ApJ...509..608T}
{Tavecchio}, F., {Maraschi}, L., \& {Ghisellini}, G. 1998, \apj, 509, 608

\bibitem[{{Timmer} \& {Koenig}(1995)}]{TimmerKoenig1995}
{Timmer}, J., \& {Koenig}, M. 1995, \aap, 300, 707

\bibitem[{{Valtonen} \& {Pihajoki}(2013)}]{Valtonen2013}
{Valtonen}, M., \& {Pihajoki}, P. 2013, \aap, 557, A28

\bibitem[{{Valtonen}(2007)}]{2007ApJ...659.1074V}
{Valtonen}, M.~J. 2007, \apj, 659, 1074

\bibitem[{{Valtonen} {et~al.}(2012){Valtonen}, {Ciprini}, \&
  {Lehto}}]{2012MNRAS.427...77V}
{Valtonen}, M.~J., {Ciprini}, S., \& {Lehto}, H.~J. 2012, \mnras, 427, 77

\bibitem[{Valtonen {et~al.}(2011)Valtonen, Mikkola, Lehto, Gopakumar, Hudec, \&
  Polednikova}]{Valtonen2011}
Valtonen, M.~J., Mikkola, S., Lehto, H.~J., {et~al.} 2011, \apj, 742, 22(12)

\bibitem[{Valtonen {et~al.}(2016)Valtonen, Zola, Ciprini, Gopakumar, Matsumoto,
  Sadakane, Kidger, Gazeas, Nilsson, Berdyugin, Piirola, Jermak, Baliyan,
  Alicavus, Boyd, Torrent, Campos, G{\'{o}}mez, Caton, Chavushyan, Dalessio,
  Debski, Dimitrov, Drozdz, Er, Erdem, P{\'{e}}rez, Ramazani, Filippenko,
  Ganesh, Garcia, Pinilla, Gopinathan, Haislip, Hudec, Hurst, Ivarsen, Jelinek,
  Joshi, Kagitani, Kaur, Keel, LaCluyze, Lee, Lindfors, de~Haro, Moore,
  Mugrauer, Nogues, Neely, Nelson, Ogloza, Okano, Pandey, Perri, Pihajoki,
  Poyner, Provencal, Pursimo, Raj, Reichart, Reinthal, Sadegi, Sakanoi,
  Gonz{\'{a}}lez, Sameer, Schweyer, Siwak, Alfaro, Sonbas, Steele, Stocke,
  Strobl, Takalo, Tomov, Espasa, Valdes, P{\'{e}}rez, Verrecchia, Webb, Yoneda,
  Zejmo, Zheng, Telting, Saario, Reynolds, Kvammen, Gafton, Karjalainen,
  Harmanen, \& Blay}]{Valtonen:2016dy}
Valtonen, M.~J., Zola, S., Ciprini, S., {et~al.} 2016, \apjl, 819, 1

\bibitem[{{Verrecchia} {et~al.}(2016){Verrecchia}, {Ciprini}, {Valtonen}, \&
  {Zola}}]{2016ATel.9709....1V}
{Verrecchia}, F., {Ciprini}, S., {Valtonen}, M., \& {Zola}, S. 2016, The
  Astronomer's Telegram, 9709

\bibitem[{Virtanen {et~al.}(2020)Virtanen, Gommers, Oliphant, Haberland, Reddy,
  Cournapeau, Burovski, Peterson, Weckesser, Bright, {van der Walt}, Brett,
  Wilson, Millman, Mayorov, Nelson, Jones, Kern, Larson, Carey, Polat, Feng,
  Moore, {VanderPlas}, Laxalde, Perktold, Cimrman, Henriksen, Quintero, Harris,
  Archibald, Ribeiro, Pedregosa, {van Mulbregt}, \& {SciPy 1.0
  Contributors}}]{2020SciPy-NMeth}
Virtanen, P., Gommers, R., Oliphant, T.~E., {et~al.} 2020, Nature Methods, 17,
  261

\bibitem[{{Wakely} \& {Horan}(2008)}]{TeVCat2008}
{Wakely}, S.~P., \& {Horan}, D. 2008, International Cosmic Ray Conference, 3,
  1341

\bibitem[{Waskom(2021)}]{Waskom2021}
Waskom, M.~L. 2021, Journal of Open Source Software, 6, 3021

\bibitem[{Wood {et~al.}(2017)Wood, Caputo, Charles, Di~Mauro, Magill, \&
  Perkins}]{Wood:2017TJ}
Wood, M., Caputo, R., Charles, E., {et~al.} 2017, PoS, ICRC2017, 824

\end{thebibliography}



\end{document}